\documentclass[journal=nalefd,manuscript=letter,layout=twocolumn]{achemso}
\usepackage[version=3]{mhchem}
\usepackage{amsmath}
\usepackage{amsfonts}
\usepackage{amssymb}
\usepackage{graphicx}
\usepackage{color, soul}

\usepackage{siunitx}
\usepackage{hyperref}
\usepackage[normalem]{ulem} 

\SectionsOn

\title[] {Heat driven transport in serial double quantum dot devices}

\author{Sven Dorsch}
\email{sven.dorsch@ftf.lth.se}
\author{Artis Svilans}
\author{Martin Josefsson}
\affiliation{Solid State Physics and NanoLund, Lund University, Box 118, SE-221 00 Lund, Sweden}
\author{Bahareh Goldozian}
\affiliation{Mathematical Physics and NanoLund, Lund University, Box 118, SE-221 00 Lund, Sweden}
\author{Mukesh Kumar}
\author{Claes Thelander}
\affiliation{Solid State Physics and NanoLund, Lund University, Box 118, SE-221 00 Lund, Sweden}
\author{Andreas Wacker}
\email{andreas.wacker@fysik.lu.se}
\affiliation{Mathematical Physics and NanoLund, Lund University, Box 118, SE-221 00 Lund, Sweden}
\author{Adam Burke}
\email{adam.burke@ftf.lth.se}
\affiliation{Solid State Physics and NanoLund, Lund University, Box 118, SE-221 00 Lund, Sweden}
\date{\today}

\begin{document}

\begin{abstract}
Studies of thermally induced transport in nanostructures provide access to an exciting regime where fluctuations are relevant, enabling the investigation of fundamental thermodynamic concepts and the realization of thermal energy harvesters. We study a serial double quantum dot formed in an InAs/InP nanowire coupled to two electron reservoirs. By means of a specially designed local metallic joule-heater, the temperature of the phonon bath in the vicinity of the double quantum dot can be enhanced. This results in phonon-assisted transport, enabling the conversion of local heat into electrical power in a nano-sized heat engine. Simultaneously, the electron temperatures of the reservoirs are affected, resulting in conventional thermoelectric transport. By detailed modelling and experimentally tuning the interdot coupling we disentangle both effects. Furthermore, we show that phonon-assisted transport gives access to the energy of excited states. Our findings demonstrate the versatility of our design to study fluctuations and fundamental nanothermodynamics.

{\bf Keywords:} nanowire, thermoelectric effect, phonon assisted transport, quantum dot, thermal energy harvesters.
\end{abstract}

\maketitle

Serial double quantum dot (DQD) devices are attractive systems for both fundamental quantum physics studies and quantum electronic-based applications with well-established charge transport properties \cite{BennettNature2000,Hanson_review, vdW_review}. Due to the electronic structure\cite{ReimannRMP2002} and controllable level detunement, DQDs are highly sensitive to their environment. Consequently, interactions of DQDs with their charge environment, photons \cite{Petta_Photon_AT_DQD, Qin_Photon_AT_PAT_DQD} and phonons \cite{Fujisawa_inel_transport_DQD, Granger_PAT_DQD_Phonon_interference, Gasser_QPC_driving_DQD, Naber_surfaceAC_PAT_DQD, Khrapai_PAT_DQD_review, Khrapai_PAT_DQD_1, Weber_conf_phonon_modes_NW_DQD, Roulleau_e_Ph_coupling_in_DQDs,Khrapai_PAT_unexplained, Hartke_Suspended_NW_DQD, hofmannarXiv2019} have been demonstrated and used for frequency resolved noise, photon and phonon detection \cite{Aguado_DQD_noise_detector,Gustavsson_Single_photon_detector_DQD,ChenSciRep2015}.

In this letter, we study the impact of interactions between a DQD and its thermal environment on transport across the device. The relation between electrical transport and temperature gradients in nanoscale systems is important for applications as energy harvesters \cite{Sothmann_thermoel_review,Sanchez_3T_harvester_theory,Thierschmann_3T_energy_harvester_nature, Jaliel_QD_energy_harvester, QD_heat_engine_fits} or fundamental studies of statistical physics\cite{HorowitzNature2020,Pietzonka_power_fluctuations,Verley_fluctuations_heat_engine,Mahan_delta_thermoelectrics, Humphrey_thermoel_nanomaterials}. In a conventional setup thermoelectricity is driven by the temperature difference created between electrical probes of a circuit which induces a current. Here, in addition to the thermoelectric effect (TE) across a DQD, we consider how a thermal reservoir decoupled from the electronic circuit of the device affects currents by phonon-assisted transport (PAT). We demonstrate how PAT and the TE can be unambiguously distinguished and used to probe excited states. Furthermore, we show that PAT can be used to harvest energy from the heated lattice.

\begin{figure*}
\includegraphics[width=7in]{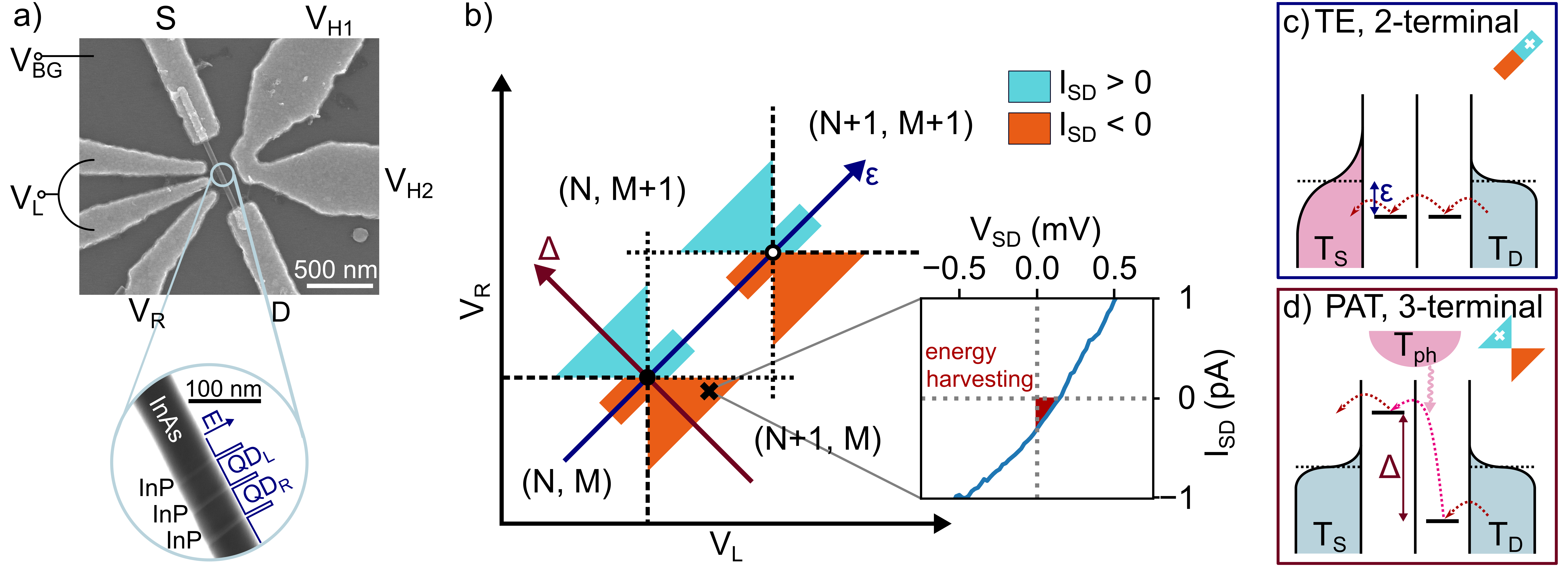} 
\caption {a) Scanning electron microscope image of the device. An InAs/InP nanowire on a substrate acting as backgate ($V_{\textrm{BG}}$) is contacted ($S$, $D$) and plunger gates ($V_{\textrm{L}}$, $V_{\textrm{R}}$) are placed in close vicinity to the DQD. The inset shows zoomed in transmission mode scanning electron microscope image of the nanowire. Three InP segments embedded in the InAs nanowire serve as tunnel barriers to epitaxially form a DQD. A heater electrode ($V_{\textrm{H1}}$, $V_{\textrm{H2}}$) is aligned with the DQD to heat the device. b) Schematic illustration of temperature gradient driven currents in the charge stability diagram of DQD devices in absence of external bias. Colored regions illustrate where currents are induced by the TE or PAT for the case $T_{\textrm{ph}}>T_{\textrm{S}}>T_{\textrm{D}}$ around the (N,M)$\to$(N+1, M+1) transition. Inset: Experimental six-time averaged $I_{\textrm{SD}}$-$V_{\textrm{SD}}$ curve, measured with a heating bias of $dV_{\textrm{H}} = 4\,\si{V}$ at a point corresponding to the black cross in the charge stability diagram. The offset of the curve towards the origin indicates that the device produces power. c) Schematic illustration of the TE, inducing transport along the energy axis $\epsilon=\mu_(-E_{\textrm{L}}+E_{\textrm{R}})/2$. d) Schematic illustration of PAT along the level detunement axis $\Delta=E_{\textrm{L}}-E_{\textrm{R}}$. The insets in (c/d) show the characteristic current patterns around the TP and the white crosses indicate the location of the level combinations shown in the panel. \label{fig1}}
\end{figure*}

Our experimental device is shown in fig. \ref{fig1}(a). A DQD is epitaxially formed within an InAs nanowire via three InP segments \cite{Fuhrer_InAs_InP_DQD, Zannier_InAs_InP_QD_growth}. Plunger gates control the energy levels of the left and right quantum dot (QD) by applying a voltage $V_{\textrm{L(R)}}$ to the left (right) plunger gate. The DQD device is coupled to the thermal reservoirs in the contacts with electron temperatures $T_{\textrm{S}}$ and $T_{\textrm{D}}$ of the source and drain contact, respectively. Additionally, the lattice acts as third thermal reservoir with a phonon temperature $T_{\textrm{ph}}$. A metallic joule-heater electrode in close vicinity to the nanowire, decoupled from the electronic system of the DQD, serves as an external heat source.

In the following, we describe the device as a DQD with weak interdot tunnel coupling $\Omega$, that is weakly coupled to the source and drain electron reservoirs. Figure \ref{fig1}(b) illustrates the charge stability diagram of the DQD in plunger gate space around a pair of triple points (TP) in the (N, M)$\to$(N+1, M+1) charge transition region. The black and white dots label the position of electron and hole TPs respectively. Here, we assume that due to substantial charging and level-quantization energies, only a single level with energy $E_\textrm{L}/E_\textrm{R}$ can be additionally occupied in the left/right dot containing already N/M electrons in this region. These addition energies are essentially linear functions of the gate voltages $E_{\textrm{L/R}}=\alpha_{\textrm{L/R}}(V_{\textrm{L/R}}^T-V_{\textrm{L/R}})$, where $V_{\textrm{L/R}}^T$ are the voltages at the electron triple point and $\alpha_{\textrm{L/R}}$ are the lever arm factors. For the sake of conveniences we define local coordinates $\epsilon=\mu-(E_\textrm{L}+E_\textrm{R})/2$ and $\Delta=E_\textrm{L}-E_\textrm{R}$, where $\mu=\mu_\textrm{S}=\mu_\textrm{D}$ is the chemical potential for vanishing source-drain bias $V_{\textrm{SD}}=0\,\si{V}$. Charging lines separate regions of different overall charge (dashed lines) and upon crossing the charge transfer line, connecting the two TPs along $\epsilon$, an electron is transferred from one QD to the other. In the absence of photons, phonons or a driving bias voltage, no net current flows and conductivity is only expected on the TPs. Cyan (orange) shaded areas indicate where heating the device via the joule-heater electrode can drive positive (negative) currents in the absence of a voltage bias.

The TE illustrated in fig. \ref{fig1}(c) occurs when $T_{\textrm{S}}\ne T_{\textrm{D}}$. The effect manifests along the axis $\epsilon$ when levels in each QD are aligned and moved in energy together. The applied temperature gradient causes different energy distributions of the electronic reservoirs around the electrochemical potentials $\mu$. The DQD connecting the two reservoirs acts as an energy filter probing the electron population imbalance at a set energy\cite{Svilans_thermoel_QD_1}. For $\epsilon < 0$ ($\epsilon > 0$) this causes electrons to flow from hot to cold (cold to hot). Consequently a reversal of the thermoelectric current polarity is observed upon crossing each TP along $\epsilon$. This effect has been studied in single QD devices\cite{QD_heat_engine_fits, Svilans_thermoel_QD_1, Svilans_thermoel_QD_2, StaringEPL1993, DzurakPRB1997, ScheibnerPRB2007, PreteNanoLett2019} as well as in DQDs\cite{Thierschmann_thermoel_DQD}.

PAT via phonon absorption is an inelastic transport mechanism requiring a detuned two-level system. Along the energy axis $\Delta=E_{\textrm{L}}-E_{\textrm{R}}$ levels on the two QDs are energetically detuned, as illustrated for $\Delta>0$ in fig. \ref{fig1}(d). For large $\Delta$ thermoelectric currents are blocked. Electrons can then only be transported from the energetically lower, occupied, to the higher, unoccupied energy level of the DQD by absorbing a phonon from the hot phonon bath and a net current can flow if $T_{\textrm{ph}}>T_{\textrm{S}},T_{\textrm{D}}$\cite{Goldozian_PAT_QmeQ_testcase}. Reverting the sign of $\Delta$ (crossing the charge transfer line where $\Delta$=0) causes the electron transport to revert direction. Consequently, PAT yields currents of opposite polarity on opposite sides of the charge transfer line as long as the spectral distribution of the phonon bath provides sufficient phonon energies to overcome the DQD level detuning. For $|\Delta|\gg k_\textrm{B}T_{\textrm{ph}}$ this condition is no longer fulfilled and PAT is suppressed. We note that a significant difference in occupation between the initial and final state for the PAT process is required \cite{Goldozian_PAT_QmeQ_testcase}. Without external bias, this requirement is only fulfilled if the two involved states are on opposite sides of the Fermi surface. Thus, PAT active regions are limited to triangular areas within the charge stability diagram as shown in fig. \ref{fig1}(b) \cite{Granger_PAT_DQD_Phonon_interference, vdW_review}.

Within these active regions, it is possible to harvest energy from an external heat source, as demonstrated in the inset of Fig. \ref{fig1}(b). Successful operation of our device as a three-terminal energy harvester is evident by the non-zero current $I_{\textrm{SD}}$ at $V_{\textrm{SD}}=0\,\si{V}$. Considering $R\approx 1\,\si{M\Omega}$ in series with the device, as used for this measurement, we extract a power of $P=RI_{\textrm{SD}}^2\approx 0.1\,\si{aW}$ \cite{QD_heat_engine_fits, Josefsson_theory_fits}.

\begin{figure}
\includegraphics[width=3.33in]{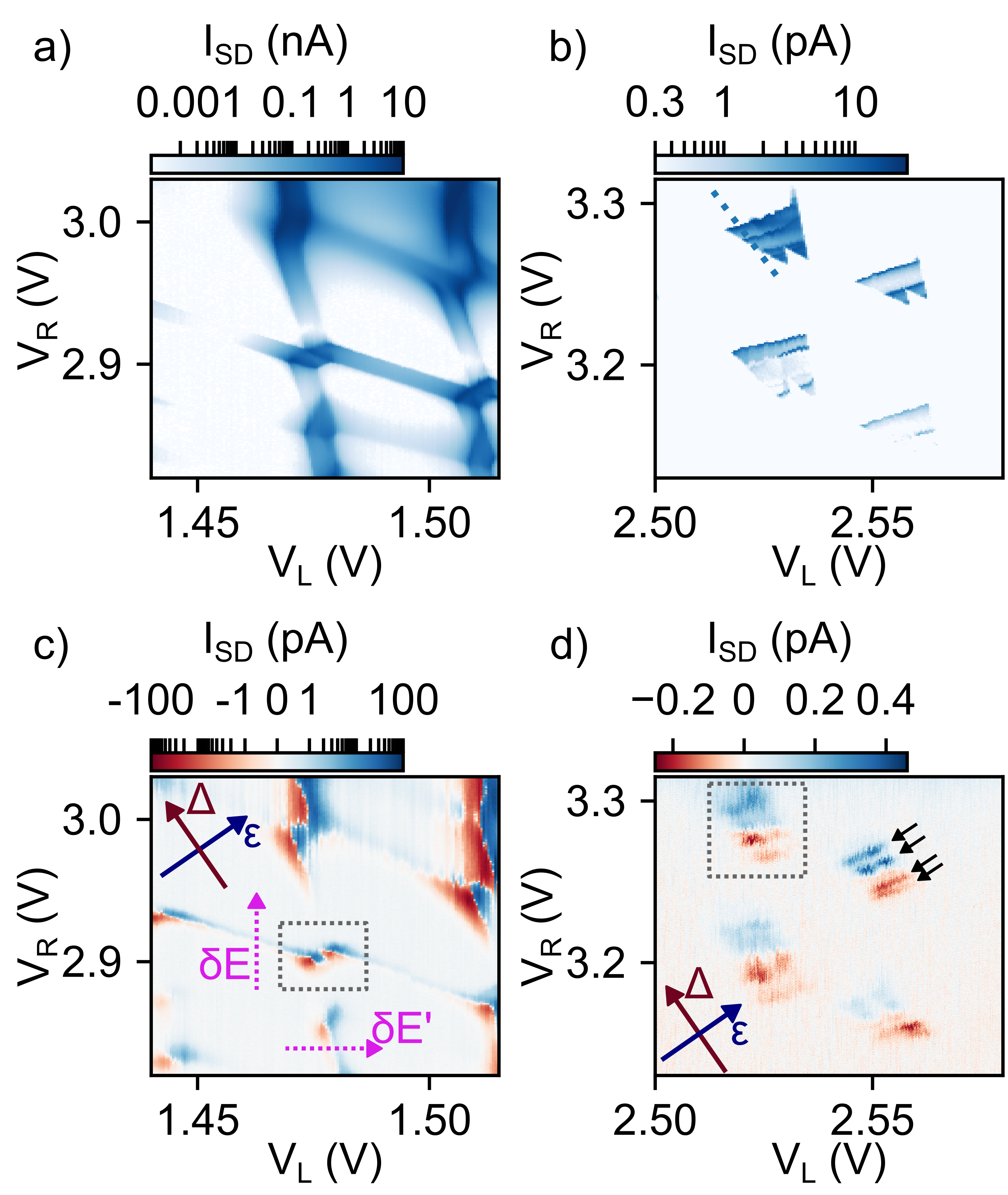} 
\caption {a) $I_{\textrm{SD}}$ is mapped as a function of the plunger gates at $V_{\textrm{H1}}=V_{\textrm{H2}}=-4\,\si{V}$ and $V_{\textrm{BG}}=0\,\si{V}$ at $V_{\textrm{SD}}=1\,\si{mV}$ in the intermediate interdot coupling regime. b) $I_{\textrm{SD}}$ is mapped as a function of the plunger gates in the weak interdot coupling regime where $V_{\textrm{BG}}=-1\,\si{V}$, $V_{\textrm{H1}}=V_{\textrm{H2}}=-4\,\si{V}$ and $V_{\textrm{SD}}=3\,\si{mV}$. c) Experimental heat driven current $I_{\textrm{SD}}$, across a serial DQD, mapped as a function of the plunger gates $V_{\textrm{L}}$ and $V_{\textrm{R}}$, measured at $V_{\textrm{SD}}=0\,\si{V}$ and $dV_{\textrm{H}}=1\,\si{V}$ in the intermediate interdot coupling regime. The current is plotted on a logarithmic scale with a linear range between $\pm 1\,\si{pA}$. d) Experimental heat driven current $I_{\textrm{SD}}$, measured at $V_{\textrm{SD}}=0\,\si{V}$ and $dV_{\textrm{H}}=4\,\si{V}$ in the weak interdot coupling regime. Dotted, grey rectangles in (c,d) label a comparable small range around a single set of TPs as in fig. \ref{fig1}(b).\label{fig2}}
\end{figure}

To experimentally study heat driven currents in our device, we first characterize the DQD via conventional finite bias spectroscopy and pick a range in the intermediate (fig. \ref{fig2}(a)) and weak (fig. \ref{fig2}(b)) interdot coupling regime. Details of the device characterization and measurement conditions are given in the supporting information (SI).

We now demonstrate a key capability of this study - by controlling the interdot tunnel coupling, we tune the dominant mechanism behind thermally driven currents between the TE and PAT. Figure \ref{fig2}(c) and (d) present exemplary measurements of purely thermally driven currents $I_{\textrm{SD}}$ across a heated DQD as a function of the plunger gate voltages $V_{\textrm{L}}$ and $V_{\textrm{R}}$ in the intermediate (c) and weak (d) interdot coupling regimes. These measurements effectively map heat driven transport signals within the DQD charge stability diagram. Dashed rectangles outline a range comparable to the schematic in Fig. \ref{fig1}(b). While maintaining $T_{\textrm{ph}}>T_{\textrm{S}}>T_{\textrm{D}}$ in both measurements, we experimentally find apparent differences in heat driven transport in the intermediate and weak interdot coupling regimes. A detailed characterization of the signals at different heating bias $dV_{\textrm{H}}=V_{\textrm{H1}}-V_{\textrm{H2}}$ is given in the SI.

In the intermediate interdot coupling regime (fig. \ref{fig2}(c)), the pronounced polarity reversal of current close to the TPs occurs along the energy axis $\epsilon$, suggesting the thermoelectric effect to be the main transport mechanism. Further, signals also appear along the charging lines and are in good agreement with thermoelectric transport across an intermediately coupled DQD by Thierschmann et al. \cite{Thierschmann_thermoel_DQD}. In contrast, in the weak interdot coupling regime (fig. \ref{fig2}(d)) currents reverse sign perpendicular to $\epsilon$, along $\Delta$, which is characteristic for a PAT dominated system. With the exception of pronounced resonances (black arrows), our results qualitatively match previous studies on phonon mediated back action of charge sensors on transport through weakly coupled DQDs \cite{Khrapai_PAT_unexplained, Khrapai_PAT_DQD_1, Khrapai_PAT_DQD_review, Granger_PAT_DQD_Phonon_interference}. 

To explain the difference between the intermediate and weak interdot coupling regime, we consider the impact of the interdot tunnel coupling $\Omega$ on heat driven transport. Discrete energy levels $E_{\textrm{L}}$ and $E_{\textrm{R}}$ confined to the left and right QD respectively only exist for $\Delta\gg\Omega$ when mixing effects are suppressed \cite{Aguado_DQD_noise_detector}. In contrast, if $\Delta\gg\Omega$ is not fulfilled, the electronic wavefunctions will extend over both QDs and mixing of $E_{\textrm{L}}$ and $E_{\textrm{R}}$ leads to the formation of bonding and antibonding molecular states $E_{\textrm{bond}}$, $E_{\textrm{antibond}}$ \cite{Blick_DQD_molecular_states}. The molecular states are energetically separated by \begin{equation}E_{\textrm{antibond}}–E_{\textrm{bond}}=[(E_{\textrm{L}}-E_{\textrm{R}})^2+4\Omega^2]^{1/2}\end{equation} and have direct consequences on both PAT and the thermoelectric effect \cite{Oosterkamp_DQD_regimes_molecules}.

For PAT, the formation of bonding and antibonding molecular states introduces new, additional constraints as phonons now have to supply a minimum energy of $E_{\textrm{ph}}>2\Omega$ even at small $\Delta$, which agrees with observations in photon assisted transport experiments \cite{Oosterkamp_DQD_regimes_molecules}. Consequently, the onset of PAT is offset towards higher phonon energies and temperatures with increasing $\Omega$. In addition, as a result of the electronic wavefunctions extending across both QDs \cite{Blick_DQD_molecular_states} the directionality of PAT is reduced as electrons now have an increased probability to tunnel back into their initial reservoir and not contribute to the detected PAT current signal. The thermoelectric effect on the other hand benefits from the formation of molecular states. While for $\Delta\gg\Omega$, elastic thermoelectric transport is blocked, for large $\Omega$ this blockade is lifted due to the extended wavefunctions and thermoelectric current contributions can be observed in an increased range along $\Delta$. Finally, along the more horizontal (vertical) charging lines where a level in the right (left) QD is aligned with $\mu_{\textrm{SD}}$ the DQD behaves as single QD with an effective left (right) tunnel coupling $\Gamma_{\textrm{L(R),eff}}\approx\Gamma_{\textrm{L(R)}}\Omega^2/\Delta^2$\cite{Gustavsson_cotunneling}. As a result of the high $\Omega$, currents are observed along these charging lines in the heated case which, along an energy axis $\delta E$ ($\delta E'$) resemble a pure TE across a single QD because no final state for a PAT process is available.

We use the observation of pure TE signals along the charging lines in the intermediate interdot coupling regime to estimate contact electron temperatures $T_{\textrm{S/D}}$ independently of the phonon temperature and find\begin{equation}\begin{split}
T_{\textrm{S}}\approx&\ 0.90\,\si{K/V}\cdot dV_{\textrm{H}}+0.13\,\si{K}\\
T_{\textrm{D}}\approx&\ 0.87\,\si{K/V}\cdot dV_{\textrm{H}}+0.13\,\si{K}\, .
\label{EqTLTRfit}
\end{split}\end{equation}
Here, the small temperature difference is in good agreement with our expectations for a nearly symmetric heating effect of the heater electrode. Details of the contact electron temperature estimation are given in the SI.

Finally, we return to weak interdot coupling and address the pronounced resonances within fig. \ref{fig2}(d) (black arrows), which were not observed by previous PAT or thermoelectric effect related studies. Comparable features are reported for experiments based on photon-assisted transport across DQDs and are attributed to multi-photon processes \cite{Oosterkamp_DQD_regimes_molecules, Petta_Photon_AT_DQD}. In contrast to those experiments, we find the resonances in fig. \ref{fig2}(d) (and SI) to be asymmetric with respect to the charge transfer line. This asymmetry, in combination with the continuous PAT signal along $\Delta$, rules out resonant one- and two-phonon processes. To study the origin of the resonances, we combine finite bias spectroscopy with heated measurements.

\begin{figure}
\includegraphics[width=3.33in]{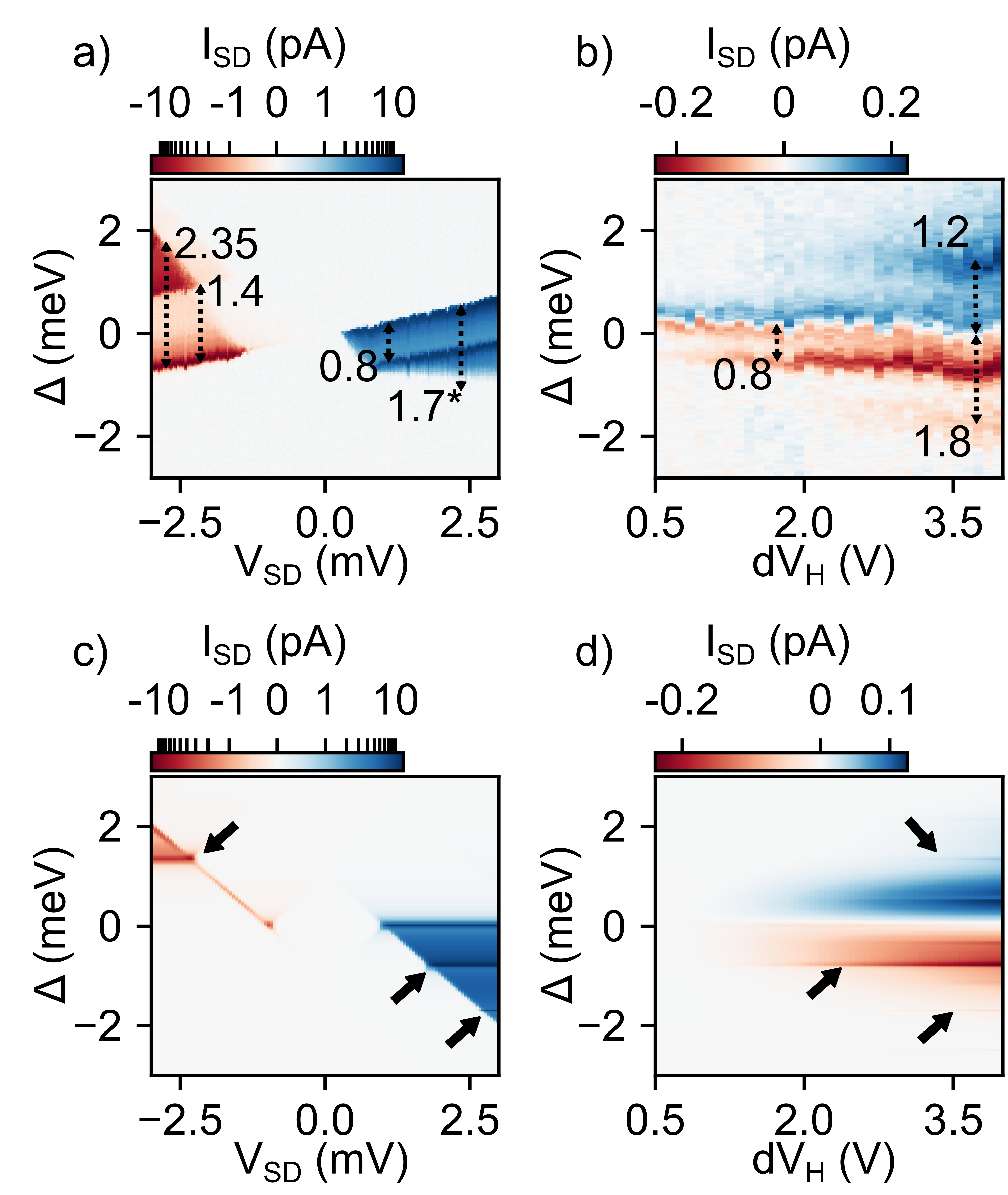} 
\caption{a) Current measurement along the cut-line $\Delta$ (blue dotted line in fig. \ref{fig2}(b)) as function of the driving bias $V_{\textrm{SD}}$ at $dV_{\textrm{H}}=0\,\si{V}$. The current is plotted logarithmic with a linear range between $\pm1\,\si{pA}$. b) Current along $\Delta$ as function of the heater voltage bias $dV_{\textrm{H}}$ at $V_{\textrm{SD}}=0\,\si{V}$. Numbers in (a) and (b) correspond to the energetic spacing from resonance to the charge transfer line in meV, indicated by dotted black arrows. The excited state at $1.7\,\si{meV}$ (labelled by * in (a)) is slightly outside the overlap of the cut-line with the bias triangle and the spacing is read out from fig. \ref{fig2}(b). Each data point in (a) and (b) is the result of two-time averaging. c) Simulation of (a) at $T_{\textrm{ph}}=T_{\textrm{S}}=T_{\textrm{D}}=130\,\si{mK}$. d) Simulation of b) using the temperature calibrations given in eqs.~(\ref{EqTLTRfit},\ref{EqTphfit}). Black arrows in (c,d) label energetically matching resonances within the bias and heat driven simulated currents.\label{fig3}}
\end{figure}

Figure \ref{fig3}(a) shows the current $I_{\textrm{SD}}$ without heating, $dV_{\textrm{H}}=0\,\si{V}$, as a function of $V_{\textrm{SD}}$ along the cut-line $\Delta$, marked by a blue dotted line in fig. \ref{fig2}(b). The cut-line is chosen such that it crosses finite bias triangles forming in both positive and negative $V_{\textrm{SD}}$ direction. We note that the observed linear slope of the groundstate (GS) to groundstate transition line (the charge transfer line), where the energetically lowest available state in each QD mediates transport, is an effect of the source contact's capacitive coupling to the DQD. $\Delta$ is manually set to zero where the charge transfer line intersects with $V_{\textrm{SD}}=0\,\si{V}$. For positive (negative) $V_{\textrm{SD}}$ we detected how the finite bias triangle grows in height with negative (positive) $\Delta$. Sudden changes in current are observed when a new excited state (ES) becomes available for transport and is initially aligned with the GS in the other QD. This allows mapping of the GS-ES spacing for the left ($\Delta < 0$) and right ($\Delta > 0$) QD. The values for the GS-ES spacings in meV are annotated in fig. \ref{fig3}(a). For positive $V_{\textrm{SD}}$, an additional GS-ES spacing of $1.7\,\si{meV}$ (labelled by *) is obtained from fig. \ref{fig2}(b). We finally note that the current suppression before the triplet state becomes available (below $1.4\,\si{meV}$) within the finite bias triangle for negative $V_{\textrm{SD}}$ is Pauli spin blockaded\cite{WeinmannPRL1995,OnoScience2002,JohnsonPRB2005}. The spin blockade is lifted along the GS-GS transition, possibly due to hyperfine- or spin-orbit-interaction \cite{Nadj_Perge_spin_block_lifted}. 

Figure \ref{fig3}(b) shows the current $I_{\textrm{SD}}$ along the cut-line $\Delta$ as a function of $dV_{\textrm{H}}$ for $V_{\textrm{SD}}=0\,\si{V}$. The PAT induced current polarity reversal at $\Delta=0$ is clearly visible. With increasing $dV_{\textrm{H}}$, the phonon temperature $T_{\textrm{ph}}$ increases, which provides larger currents. Furthermore, the PAT currents are visible for $|\Delta|$ being less than a few $k_\textrm{B}T_{\textrm{ph}}$, as phonons with higher energies are entirely frozen out. Thus the PAT active regions grow with increasing $dV_{\textrm{H}}$. Resonances, as observed in fig. \ref{fig2}(d), begin to appear with increasing heating voltages $dV_{\textrm{H}}$. The energy spacing in meV from resonance to charge transfer line is annotated in fig. \ref{fig3}(b) and we find similar values to the GS-ES spacings in fig. \ref{fig3}(a). 

One possible cause for the resonances is a TE through aligned GS-ES configurations. This would lead to a polarity reversal of the current on the resonances once the aligned GS-ES levels are energetically pulled below $\mu$ as we follow the resonance in direction of $\epsilon$. We interpret the absence of this polarity reversal in fig. \ref{fig2}(d) as an indication of PAT through excited states as origin for the resonances.

In order to verify PAT as the origin of the observed features, we performed detailed simulations of the transport through a DQD system with two spin degenerate levels in each QD (ground levels $E_{\textrm{L/R}}$ and excited levels with spacing $\Delta E_{L}=0.8 \si{meV}$ and  $\Delta E_{R}=1.7 \si{meV}$). Furthermore, we use the Coulomb interaction terms for interdot $U=8\,\textrm{meV}$, for intradot $U_n=1.4 \si{meV}$, and for intradot exchange $U_\textrm{ex}=0.4 \si{meV}$. These parameters are  motivated in the SI, where additional simulation details are given. All transport simulations are done with the open source simulation tool QMEQ [for \textit{Quantum Master Equation for Quantum dot transport
calculations}] \cite{QmeQ} with the recent inclusion of phonon scattering \cite{Goldozian_PAT_QmeQ_testcase}.

\begin{figure}
\includegraphics[width=3.33in]{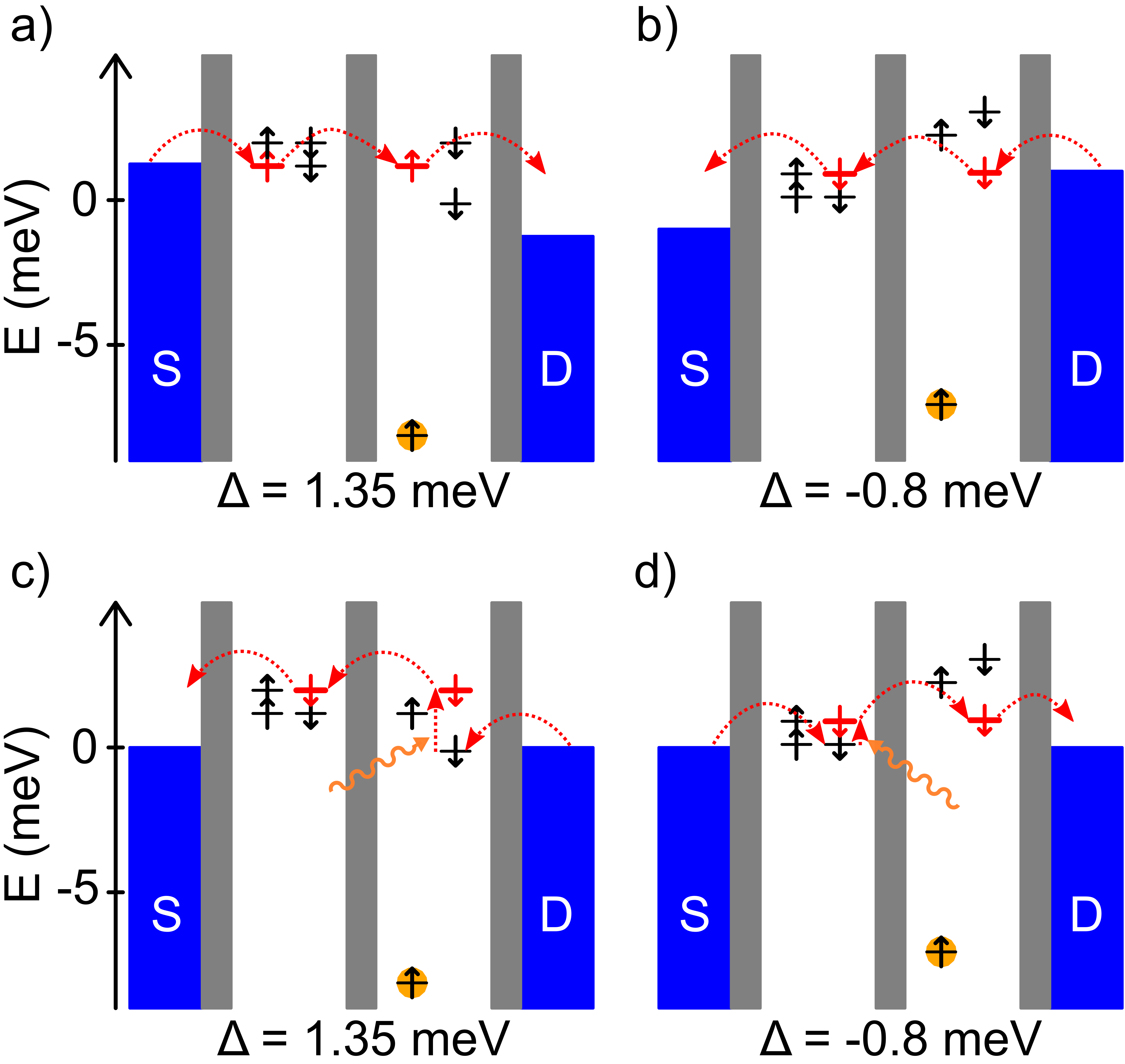} 
\caption{Electrical- (a/b) and thermal- (c/d) bias driven transport mechanisms along the cut-lines shown in fig. \ref{fig3}(c/d). Addition energies for the state with a single electron in the right QD (marked by a filled orange circle) are shown. (a/c) schematically show transport via bias/PAT at the $\Delta=1.35\,\si{meV}$ resonance. (b/d) schematically show transport via bias/PAT at the $\Delta=-0.8\,\si{meV}$ resonance. Red dotted arrows highlight the relevant transport mechanisms and red levels indicate the involved level resonances.\label{figLevels}}
\end{figure}

In the following, we focus on the $(0,1)\to(1,2)$ charge transition region (where the left/right number denotes the occupation of the left/right QD). Figure \ref{fig3}(c) shows the simulated current $I_{\textrm{SD}}$ as function of $V_{\textrm{SD}}$ without heating in the weak interdot coupling regime along a comparable cut-line $\Delta=E_\textrm{L}-E_\textrm{R}-\Delta_0$ to the one illustrated by the dotted blue line in fig. \ref{fig2}(b). Here, the ground states  GS$(1,1)$ and GS$(0,2)$ are in resonance for $E_\textrm{L}=E_\textrm{R}+\Delta_0$. For negative bias (electrons flowing to the right) we see the common spin blockade for the  triplet configuration of GS$(1,1)$ \cite{WeinmannPRL1995,OnoScience2002,JohnsonPRB2005}. The spin blockade is lifted when the triplet ES$(0,2_e)$ gets in resonance with GS$(1,1)$ for $\Delta=\sqrt{\Delta E_{\textrm{R}}^2+U_\textrm{ex}^2}-U_\textrm{ex}\approx 1.35\,\si{meV}$, see left black arrow. Figure \ref{figLevels}(a) displays this transport mechanism and the behaviour is comparable to the data in fig. \ref{fig3}(a), where, however, the spin blockade is not complete. Similarly, fig. \ref{figLevels}(b) shows the level position for $\Delta\approx -\Delta E_{\textrm{L}}=-0.8\,\si{\meV}$ (ES$(1_e,1)$ is in resonance with GS$(0,2)$), where in good agreement with the experiment a resonance is observed for positive bias in fig. \ref{fig3}(c). In addition, we find a weak resonance at $\Delta\approx -1.7\,\si{meV}$, where the singlets ES$(1,1_e)$ and GS$(0,2)$ are in resonance. This coupling is a many-body effect, which is not present in the dominant single-particle product states. The experimental data shows resonances at comparable values of $\Delta$, indicating the accuracy of the extracted parameters. Simulation results for the charging diagrams corresponding to fig. \ref{fig2} are presented in the SI.

To model the side-heating, we assume the phonon temperature
\begin{equation}
T_\textrm{ph}=1.2\,\si{K/V}\cdot dV_{\textrm{H}}+0.13\,\si{K}\, ,\label{EqTphfit}
\end{equation}
as the heating bias $dV_{\textrm{H}}$ should have a stronger impact on the phonons in the DQD than on the electrons in the leads, compare eq.~(\ref{EqTLTRfit}), which is simultaneously used here. As can be seen in fig. \ref{fig3}(d), the use of the phonon and electron temperature dependencies in eqs.~(\ref{EqTLTRfit},\ref{EqTphfit}) well reproduce the experimental data in fig. \ref{fig3}(b). The magnitude of the currents for negative $\Delta$ matches well with the experiment, and is however slightly smaller for positive $\Delta$. 

With increasing $dV_{\textrm{H}}$, several resonances  (labelled by black arrows) occur for specific values of $\Delta$, see fig. \ref{fig3}(d), which are also present in the experimental data. These peaks can be attributed to the alignment of different excitations. While the typical PAT signal is the result of electrons lifted into virtual states via phonon absorption \cite{Qin_Photon_AT_PAT_DQD}, followed by tunnelling into a real state in the second dot, we expect an increased probability for PAT if only real states are involved. This is illustrated in fig. \ref{figLevels}(d). For the resonance between GS$(0,2)$ and ES$(1_e,1)$  for $\Delta=-0.8\,\si{meV}$, heated phonons provide the excitation from GS$(1,1)$ to ES$(1_e,1)$, with subsequent tunnelling to GS$(0,2)$, resulting in a negative current.

The PAT situation is more complex for the resonance between the triplets
GS$(1,1)$ and ES$(0,2_e)$ at $\Delta\approx 1.35\,\si{meV}$ sketched in  fig. \ref{figLevels}(a). The corresponding PAT process requires several preceding intermediate steps to occupy the triplet ES$(0,2_e)$. In addition, due to our particular choice $\Delta E_{\textrm{R}}=2U_\textrm{ex}$, there is a second resonance between the singlets ES$(1_e,1)$ and ES$(0,2_e)$, see fig. \ref{figLevels}(c). 
For heated phonons,  GS$(0,2)$ is excited to singlet 
ES$(0,2_e)$, providing a distinct PAT resonance with positive current.  
The presence of different resonances may explain the relatively wide experimental peak around $\Delta\approx 1.2 \si{meV}$ in fig. \ref{fig3}(b). 

We note that the wide peak around  $\Delta\approx 0.6\,\si{\meV}$ in fig. \ref{fig3} (d) is related to the phonon absorption rate, as $J(E)/(e^{E/kT_\textrm{ph}}-1)$, where $J(E)$ is the phonon spectral density, has a maximum at $E=0.6\,\si{\meV}$ for $dV_{\textrm{H}}=4\,\si{\V}$. Changing the phenomenological extent of the wavefunctions we can change the location of this peak and the extension of currents upon variations of $\Delta$.

In summary, we present a detailed, combined experimental and theoretical study of heat driven transport in a DQD coupled to three thermal reservoirs. We show how via tuning the interdot tunnel coupling, the dominant transport mechanism can be tuned between the two-terminal TE and three-terminal PAT, which is further found to be sensitive to excited states. Consequently, we demonstrate the applicability of conventional DQD devices as highly versatile and controllable energy harvesters and provide a platform and the tools to conduct and analyze fundamental nanothermodynamic experiments.

\begin{acknowledgement}
The authors thank H. Linke for fruitful discussions and acknowledge funding by the Knut and Alice Wallenberg Foundation (KAW) (project2016.0089), the Swedish Research Council (VR) (project 2015-00619), the Marie Skłodowska Curie Actions, Cofund, Project INCA 600398 and by NanoLund. Device fabrication was carried out in the Lund Nano Lab (LNL) and computational resources were provided by the Swedish National Infrastructure for Computing (SNIC) at LUNARC.
\end{acknowledgement}

\bibliography{PAT_Bib}

\providecommand{\latin}[1]{#1}
\makeatletter
\providecommand{\doi}
  {\begingroup\let\do\@makeother\dospecials
  \catcode`\{=1 \catcode`\}=2 \doi@aux}
\providecommand{\doi@aux}[1]{\endgroup\texttt{#1}}
\makeatother
\providecommand*\mcitethebibliography{\thebibliography}
\csname @ifundefined\endcsname{endmcitethebibliography}
  {\let\endmcitethebibliography\endthebibliography}{}
\begin{mcitethebibliography}{50}
\providecommand*\natexlab[1]{#1}
\providecommand*\mciteSetBstSublistMode[1]{}
\providecommand*\mciteSetBstMaxWidthForm[2]{}
\providecommand*\mciteBstWouldAddEndPuncttrue
  {\def\EndOfBibitem{\unskip.}}
\providecommand*\mciteBstWouldAddEndPunctfalse
  {\let\EndOfBibitem\relax}
\providecommand*\mciteSetBstMidEndSepPunct[3]{}
\providecommand*\mciteSetBstSublistLabelBeginEnd[3]{}
\providecommand*\EndOfBibitem{}
\mciteSetBstSublistMode{f}
\mciteSetBstMaxWidthForm{subitem}{(\alph{mcitesubitemcount})}
\mciteSetBstSublistLabelBeginEnd
  {\mcitemaxwidthsubitemform\space}
  {\relax}
  {\relax}

\bibitem[Bennett and DiVincenzo(2000)Bennett, and
  DiVincenzo]{BennettNature2000}
Bennett,~C.~H.; DiVincenzo,~D.~P. Quantum information and computation.
  \emph{Nature} \textbf{2000}, \emph{404}, 247--255\relax
\mciteBstWouldAddEndPuncttrue
\mciteSetBstMidEndSepPunct{\mcitedefaultmidpunct}
{\mcitedefaultendpunct}{\mcitedefaultseppunct}\relax
\EndOfBibitem
\bibitem[Hanson \latin{et~al.}(2007)Hanson, Kouwenhoven, Petta, Tarucha, and
  Vandersypen]{Hanson_review}
Hanson,~R.; Kouwenhoven,~L.~P.; Petta,~J.~R.; Tarucha,~S.; Vandersypen,~L.
  M.~K. Spins in few-electron quantum dots. \emph{Rev. Mod. Phys.}
  \textbf{2007}, \emph{79}, 1217--1265\relax
\mciteBstWouldAddEndPuncttrue
\mciteSetBstMidEndSepPunct{\mcitedefaultmidpunct}
{\mcitedefaultendpunct}{\mcitedefaultseppunct}\relax
\EndOfBibitem
\bibitem[van~der Wiel \latin{et~al.}(2002)van~der Wiel, De~Franceschi,
  Elzerman, Fujisawa, Tarucha, and Kouwenhoven]{vdW_review}
van~der Wiel,~W.~G.; De~Franceschi,~S.; Elzerman,~J.~M.; Fujisawa,~T.;
  Tarucha,~S.; Kouwenhoven,~L.~P. Electron transport through double quantum
  dots. \emph{Rev. Mod. Phys.} \textbf{2002}, \emph{75}, 1--22\relax
\mciteBstWouldAddEndPuncttrue
\mciteSetBstMidEndSepPunct{\mcitedefaultmidpunct}
{\mcitedefaultendpunct}{\mcitedefaultseppunct}\relax
\EndOfBibitem
\bibitem[Reimann and Manninen(2002)Reimann, and Manninen]{ReimannRMP2002}
Reimann,~S.~M.; Manninen,~M. Electronic structure of quantum dots. \emph{Rev.
  Mod. Phys.} \textbf{2002}, \emph{74}, 1283--1342\relax
\mciteBstWouldAddEndPuncttrue
\mciteSetBstMidEndSepPunct{\mcitedefaultmidpunct}
{\mcitedefaultendpunct}{\mcitedefaultseppunct}\relax
\EndOfBibitem
\bibitem[Petta \latin{et~al.}(2004)Petta, Johnson, Marcus, Hanson, and
  Gossard]{Petta_Photon_AT_DQD}
Petta,~J.~R.; Johnson,~A.~C.; Marcus,~C.~M.; Hanson,~M.~P.; Gossard,~A.~C.
  Manipulation of a Single Charge in a Double Quantum Dot. \emph{Phys. Rev.
  Lett.} \textbf{2004}, \emph{93}, 186802\relax
\mciteBstWouldAddEndPuncttrue
\mciteSetBstMidEndSepPunct{\mcitedefaultmidpunct}
{\mcitedefaultendpunct}{\mcitedefaultseppunct}\relax
\EndOfBibitem
\bibitem[Qin \latin{et~al.}(2001)Qin, Holleitner, Eberl, and
  Blick]{Qin_Photon_AT_PAT_DQD}
Qin,~H.; Holleitner,~A.~W.; Eberl,~K.; Blick,~R.~H. Coherent superposition of
  photon- and phonon-assisted tunneling in coupled quantum dots. \emph{Phys.
  Rev. B} \textbf{2001}, \emph{64}, 241302\relax
\mciteBstWouldAddEndPuncttrue
\mciteSetBstMidEndSepPunct{\mcitedefaultmidpunct}
{\mcitedefaultendpunct}{\mcitedefaultseppunct}\relax
\EndOfBibitem
\bibitem[Fujisawa \latin{et~al.}(1998)Fujisawa, Oosterkamp, van~der Wiel,
  Broer, Aguado, Tarucha, and Kouwenhoven]{Fujisawa_inel_transport_DQD}
Fujisawa,~T.; Oosterkamp,~T.~H.; van~der Wiel,~W.~G.; Broer,~B.~W.; Aguado,~R.;
  Tarucha,~S.; Kouwenhoven,~L.~P. Spontaneous Emission Spectrum in Double
  Quantum Dot Devices. \emph{Science} \textbf{1998}, \emph{282}, 932--935\relax
\mciteBstWouldAddEndPuncttrue
\mciteSetBstMidEndSepPunct{\mcitedefaultmidpunct}
{\mcitedefaultendpunct}{\mcitedefaultseppunct}\relax
\EndOfBibitem
\bibitem[Granger \latin{et~al.}(2012)Granger, Taubert, Young, Gaudreau, Kam,
  Studenikin, Zawadzki, Harbusch, Schuh, Wegscheider, Wasilewski, Clerk,
  Ludwig, and Sachrajda]{Granger_PAT_DQD_Phonon_interference}
Granger,~G.; Taubert,~D.; Young,~C.~E.; Gaudreau,~L.; Kam,~A.;
  Studenikin,~S.~A.; Zawadzki,~P.; Harbusch,~D.; Schuh,~D.; Wegscheider,~W.;
  Wasilewski,~Z.~R.; Clerk,~A.~A.; Ludwig,~S.; Sachrajda,~A.~S. Quantum
  interference and phonon-mediated back-action in lateral quantum-dot circuits.
  \emph{Nature Physics} \textbf{2012}, \emph{8}, 522--527\relax
\mciteBstWouldAddEndPuncttrue
\mciteSetBstMidEndSepPunct{\mcitedefaultmidpunct}
{\mcitedefaultendpunct}{\mcitedefaultseppunct}\relax
\EndOfBibitem
\bibitem[Gasser \latin{et~al.}(2009)Gasser, Gustavsson, K\"ung, Ensslin, Ihn,
  Driscoll, and Gossard]{Gasser_QPC_driving_DQD}
Gasser,~U.; Gustavsson,~S.; K\"ung,~B.; Ensslin,~K.; Ihn,~T.; Driscoll,~D.~C.;
  Gossard,~A.~C. Statistical electron excitation in a double quantum dot
  induced by two independent quantum point contacts. \emph{Phys. Rev. B}
  \textbf{2009}, \emph{79}, 035303\relax
\mciteBstWouldAddEndPuncttrue
\mciteSetBstMidEndSepPunct{\mcitedefaultmidpunct}
{\mcitedefaultendpunct}{\mcitedefaultseppunct}\relax
\EndOfBibitem
\bibitem[Naber \latin{et~al.}(2006)Naber, Fujisawa, Liu, and van~der
  Wiel]{Naber_surfaceAC_PAT_DQD}
Naber,~W. J.~M.; Fujisawa,~T.; Liu,~H.~W.; van~der Wiel,~W.~G.
  Surface-Acoustic-Wave-Induced Transport in a Double Quantum Dot. \emph{Phys.
  Rev. Lett.} \textbf{2006}, \emph{96}, 136807\relax
\mciteBstWouldAddEndPuncttrue
\mciteSetBstMidEndSepPunct{\mcitedefaultmidpunct}
{\mcitedefaultendpunct}{\mcitedefaultseppunct}\relax
\EndOfBibitem
\bibitem[Khrapai \latin{et~al.}(2008)Khrapai, Ludwig, Kotthaus, Tranitz, and
  Wegscheider]{Khrapai_PAT_DQD_review}
Khrapai,~V.~S.; Ludwig,~S.; Kotthaus,~J.~P.; Tranitz,~H.~P.; Wegscheider,~W.
  Nonequilibrium interactions between two quantum circuits. \emph{Journal of
  Physics: Condensed Matter} \textbf{2008}, \emph{20}, 454205\relax
\mciteBstWouldAddEndPuncttrue
\mciteSetBstMidEndSepPunct{\mcitedefaultmidpunct}
{\mcitedefaultendpunct}{\mcitedefaultseppunct}\relax
\EndOfBibitem
\bibitem[Khrapai \latin{et~al.}(2008)Khrapai, Ludwig, Kotthaus, Tranitz, and
  Wegscheider]{Khrapai_PAT_DQD_1}
Khrapai,~V.; Ludwig,~S.; Kotthaus,~J.; Tranitz,~H.; Wegscheider,~W.
  Nonequilibrium phenomena in adjacent electrically isolated nanostructures.
  \emph{Physica E: Low-dimensional Systems and Nanostructures} \textbf{2008},
  \emph{40}, 995--998\relax
\mciteBstWouldAddEndPuncttrue
\mciteSetBstMidEndSepPunct{\mcitedefaultmidpunct}
{\mcitedefaultendpunct}{\mcitedefaultseppunct}\relax
\EndOfBibitem
\bibitem[Weber \latin{et~al.}(2010)Weber, Fuhrer, Fasth, Lindwall, Samuelson,
  and Wacker]{Weber_conf_phonon_modes_NW_DQD}
Weber,~C.; Fuhrer,~A.; Fasth,~C.; Lindwall,~G.; Samuelson,~L.; Wacker,~A.
  Probing Confined Phonon Modes by Transport through a Nanowire Double Quantum
  Dot. \emph{Phys. Rev. Lett.} \textbf{2010}, \emph{104}, 036801\relax
\mciteBstWouldAddEndPuncttrue
\mciteSetBstMidEndSepPunct{\mcitedefaultmidpunct}
{\mcitedefaultendpunct}{\mcitedefaultseppunct}\relax
\EndOfBibitem
\bibitem[Roulleau \latin{et~al.}(2011)Roulleau, Baer, Choi, Molitor,
  G{\"u}ttinger, M{\"u}ller, Dr{\"o}scher, Ensslin, and
  Ihn]{Roulleau_e_Ph_coupling_in_DQDs}
Roulleau,~P.; Baer,~S.; Choi,~T.; Molitor,~F.; G{\"u}ttinger,~J.;
  M{\"u}ller,~T.; Dr{\"o}scher,~S.; Ensslin,~K.; Ihn,~T. Coherent
  electron-phonon coupling in tailored quantum systems. \emph{Nature
  Communications} \textbf{2011}, \emph{2}, 239\relax
\mciteBstWouldAddEndPuncttrue
\mciteSetBstMidEndSepPunct{\mcitedefaultmidpunct}
{\mcitedefaultendpunct}{\mcitedefaultseppunct}\relax
\EndOfBibitem
\bibitem[Khrapai \latin{et~al.}(2006)Khrapai, Ludwig, Kotthaus, Tranitz, and
  Wegscheider]{Khrapai_PAT_unexplained}
Khrapai,~V.~S.; Ludwig,~S.; Kotthaus,~J.~P.; Tranitz,~H.~P.; Wegscheider,~W.
  Double-Dot Quantum Ratchet Driven by an Independently Biased Quantum Point
  Contact. \emph{Phys. Rev. Lett.} \textbf{2006}, \emph{97}, 176803\relax
\mciteBstWouldAddEndPuncttrue
\mciteSetBstMidEndSepPunct{\mcitedefaultmidpunct}
{\mcitedefaultendpunct}{\mcitedefaultseppunct}\relax
\EndOfBibitem
\bibitem[Hartke \latin{et~al.}(2018)Hartke, Liu, Gullans, and
  Petta]{Hartke_Suspended_NW_DQD}
Hartke,~T.~R.; Liu,~Y.-Y.; Gullans,~M.~J.; Petta,~J.~R. Microwave Detection of
  Electron-Phonon Interactions in a Cavity-Coupled Double Quantum Dot.
  \emph{Phys. Rev. Lett.} \textbf{2018}, \emph{120}, 097701\relax
\mciteBstWouldAddEndPuncttrue
\mciteSetBstMidEndSepPunct{\mcitedefaultmidpunct}
{\mcitedefaultendpunct}{\mcitedefaultseppunct}\relax
\EndOfBibitem
\bibitem[Hofmann \latin{et~al.}(2019)Hofmann, Karlewski, Heimes, Reichl,
  Wegscheider, Sch{\"o}n, Ensslin, Ihn, and Maisi]{hofmannarXiv2019}
Hofmann,~A.; Karlewski,~C.; Heimes,~A.; Reichl,~C.; Wegscheider,~W.;
  Sch{\"o}n,~G.; Ensslin,~K.; Ihn,~T.; Maisi,~V.~F. Phonon spectral density in
  a GaAs/AlGaAs double quantum dot. \emph{arXiv preprint arXiv:1912.10696}
  \textbf{2019}, \relax
\mciteBstWouldAddEndPunctfalse
\mciteSetBstMidEndSepPunct{\mcitedefaultmidpunct}
{}{\mcitedefaultseppunct}\relax
\EndOfBibitem
\bibitem[Aguado and Kouwenhoven(2000)Aguado, and
  Kouwenhoven]{Aguado_DQD_noise_detector}
Aguado,~R.; Kouwenhoven,~L.~P. Double Quantum Dots as Detectors of
  High-Frequency Quantum Noise in Mesoscopic Conductors. \emph{Phys. Rev.
  Lett.} \textbf{2000}, \emph{84}, 1986--1989\relax
\mciteBstWouldAddEndPuncttrue
\mciteSetBstMidEndSepPunct{\mcitedefaultmidpunct}
{\mcitedefaultendpunct}{\mcitedefaultseppunct}\relax
\EndOfBibitem
\bibitem[Gustavsson \latin{et~al.}(2007)Gustavsson, Studer, Leturcq, Ihn,
  Ensslin, Driscoll, and Gossard]{Gustavsson_Single_photon_detector_DQD}
Gustavsson,~S.; Studer,~M.; Leturcq,~R.; Ihn,~T.; Ensslin,~K.; Driscoll,~D.~C.;
  Gossard,~A.~C. Frequency-Selective Single-Photon Detection Using a Double
  Quantum Dot. \emph{Phys. Rev. Lett.} \textbf{2007}, \emph{99}, 206804\relax
\mciteBstWouldAddEndPuncttrue
\mciteSetBstMidEndSepPunct{\mcitedefaultmidpunct}
{\mcitedefaultendpunct}{\mcitedefaultseppunct}\relax
\EndOfBibitem
\bibitem[Chen \latin{et~al.}(2015)Chen, Sato, Kosaka, Hashisaka, Muraki, and
  Fujisawa]{ChenSciRep2015}
Chen,~J. C.~H.; Sato,~Y.; Kosaka,~R.; Hashisaka,~M.; Muraki,~K.; Fujisawa,~T.
  Enhanced electron-phonon coupling for a semiconductor charge qubit in a
  surface phonon cavity. \emph{Scientific Reports} \textbf{2015}, \emph{5},
  15176\relax
\mciteBstWouldAddEndPuncttrue
\mciteSetBstMidEndSepPunct{\mcitedefaultmidpunct}
{\mcitedefaultendpunct}{\mcitedefaultseppunct}\relax
\EndOfBibitem
\bibitem[Sothmann \latin{et~al.}(2014)Sothmann, S{\'{a}}nchez, and
  Jordan]{Sothmann_thermoel_review}
Sothmann,~B.; S{\'{a}}nchez,~R.; Jordan,~A.~N. Thermoelectric energy harvesting
  with quantum dots. \emph{Nanotechnology} \textbf{2014}, \emph{26},
  032001\relax
\mciteBstWouldAddEndPuncttrue
\mciteSetBstMidEndSepPunct{\mcitedefaultmidpunct}
{\mcitedefaultendpunct}{\mcitedefaultseppunct}\relax
\EndOfBibitem
\bibitem[S\'anchez and B\"uttiker(2011)S\'anchez, and
  B\"uttiker]{Sanchez_3T_harvester_theory}
S\'anchez,~R.; B\"uttiker,~M. Optimal energy quanta to current conversion.
  \emph{Phys. Rev. B} \textbf{2011}, \emph{83}, 085428\relax
\mciteBstWouldAddEndPuncttrue
\mciteSetBstMidEndSepPunct{\mcitedefaultmidpunct}
{\mcitedefaultendpunct}{\mcitedefaultseppunct}\relax
\EndOfBibitem
\bibitem[Thierschmann \latin{et~al.}(2015)Thierschmann, S{\'a}nchez, Sothmann,
  Arnold, Heyn, Hansen, Buhmann, and
  Molenkamp]{Thierschmann_3T_energy_harvester_nature}
Thierschmann,~H.; S{\'a}nchez,~R.; Sothmann,~B.; Arnold,~F.; Heyn,~C.;
  Hansen,~W.; Buhmann,~H.; Molenkamp,~L.~W. Three-terminal energy harvester
  with coupled quantum dots. \emph{Nature Nanotechnology} \textbf{2015},
  \emph{10}, 854 EP --\relax
\mciteBstWouldAddEndPuncttrue
\mciteSetBstMidEndSepPunct{\mcitedefaultmidpunct}
{\mcitedefaultendpunct}{\mcitedefaultseppunct}\relax
\EndOfBibitem
\bibitem[Jaliel \latin{et~al.}(2019)Jaliel, Puddy, S\'anchez, Jordan, Sothmann,
  Farrer, Griffiths, Ritchie, and Smith]{Jaliel_QD_energy_harvester}
Jaliel,~G.; Puddy,~R.~K.; S\'anchez,~R.; Jordan,~A.~N.; Sothmann,~B.;
  Farrer,~I.; Griffiths,~J.~P.; Ritchie,~D.~A.; Smith,~C.~G. Experimental
  Realization of a Quantum Dot Energy Harvester. \emph{Phys. Rev. Lett.}
  \textbf{2019}, \emph{123}, 117701\relax
\mciteBstWouldAddEndPuncttrue
\mciteSetBstMidEndSepPunct{\mcitedefaultmidpunct}
{\mcitedefaultendpunct}{\mcitedefaultseppunct}\relax
\EndOfBibitem
\bibitem[Josefsson \latin{et~al.}(2018)Josefsson, Svilans, Burke, Hoffmann,
  Fahlvik, Thelander, Leijnse, and Linke]{QD_heat_engine_fits}
Josefsson,~M.; Svilans,~A.; Burke,~A.; Hoffmann,~E.; Fahlvik,~S.;
  Thelander,~C.; Leijnse,~M.; Linke,~H. A quantum-dot heat engine operating
  close to the thermodynamic efficiency limits. \emph{Nature Nanotechnology}
  \textbf{2018}, \emph{13}, 920--924\relax
\mciteBstWouldAddEndPuncttrue
\mciteSetBstMidEndSepPunct{\mcitedefaultmidpunct}
{\mcitedefaultendpunct}{\mcitedefaultseppunct}\relax
\EndOfBibitem
\bibitem[Horowitz and Gingrich(2020)Horowitz, and Gingrich]{HorowitzNature2020}
Horowitz,~J.~M.; Gingrich,~T.~R. Thermodynamic uncertainty relations constrain
  non-equilibrium fluctuations. \emph{Nature Physics} \textbf{2020}, \emph{16},
  15--20\relax
\mciteBstWouldAddEndPuncttrue
\mciteSetBstMidEndSepPunct{\mcitedefaultmidpunct}
{\mcitedefaultendpunct}{\mcitedefaultseppunct}\relax
\EndOfBibitem
\bibitem[Pietzonka and Seifert(2018)Pietzonka, and
  Seifert]{Pietzonka_power_fluctuations}
Pietzonka,~P.; Seifert,~U. Universal Trade-Off between Power, Efficiency, and
  Constancy in Steady-State Heat Engines. \emph{Phys. Rev. Lett.}
  \textbf{2018}, \emph{120}, 190602\relax
\mciteBstWouldAddEndPuncttrue
\mciteSetBstMidEndSepPunct{\mcitedefaultmidpunct}
{\mcitedefaultendpunct}{\mcitedefaultseppunct}\relax
\EndOfBibitem
\bibitem[Verley \latin{et~al.}(2014)Verley, Esposito, Willaert, and Van~den
  Broeck]{Verley_fluctuations_heat_engine}
Verley,~G.; Esposito,~M.; Willaert,~T.; Van~den Broeck,~C. The unlikely Carnot
  efficiency. \emph{Nature Communications} \textbf{2014}, \emph{5}, 4721\relax
\mciteBstWouldAddEndPuncttrue
\mciteSetBstMidEndSepPunct{\mcitedefaultmidpunct}
{\mcitedefaultendpunct}{\mcitedefaultseppunct}\relax
\EndOfBibitem
\bibitem[Mahan and Sofo(1996)Mahan, and Sofo]{Mahan_delta_thermoelectrics}
Mahan,~G.~D.; Sofo,~J.~O. The best thermoelectric. \emph{Proceedings of the
  National Academy of Sciences} \textbf{1996}, \emph{93}, 7436--7439\relax
\mciteBstWouldAddEndPuncttrue
\mciteSetBstMidEndSepPunct{\mcitedefaultmidpunct}
{\mcitedefaultendpunct}{\mcitedefaultseppunct}\relax
\EndOfBibitem
\bibitem[Humphrey and Linke(2005)Humphrey, and
  Linke]{Humphrey_thermoel_nanomaterials}
Humphrey,~T.~E.; Linke,~H. Reversible Thermoelectric Nanomaterials. \emph{Phys.
  Rev. Lett.} \textbf{2005}, \emph{94}, 096601\relax
\mciteBstWouldAddEndPuncttrue
\mciteSetBstMidEndSepPunct{\mcitedefaultmidpunct}
{\mcitedefaultendpunct}{\mcitedefaultseppunct}\relax
\EndOfBibitem
\bibitem[Fuhrer \latin{et~al.}(2007)Fuhrer, Fröberg, Pedersen, Larsson,
  Wacker, Pistol, and Samuelson]{Fuhrer_InAs_InP_DQD}
Fuhrer,~A.; Fröberg,~L.~E.; Pedersen,~J.~N.; Larsson,~M.~W.; Wacker,~A.;
  Pistol,~M.-E.; Samuelson,~L. Few Electron Double Quantum Dots in InAs/InP
  Nanowire Heterostructures. \emph{Nano Lett.} \textbf{2007}, \emph{7},
  243\relax
\mciteBstWouldAddEndPuncttrue
\mciteSetBstMidEndSepPunct{\mcitedefaultmidpunct}
{\mcitedefaultendpunct}{\mcitedefaultseppunct}\relax
\EndOfBibitem
\bibitem[Zannier \latin{et~al.}(2019)Zannier, Rossi, Ercolani, and
  Sorba]{Zannier_InAs_InP_QD_growth}
Zannier,~V.; Rossi,~F.; Ercolani,~D.; Sorba,~L. Growth dynamics of {InAs}/{InP}
  nanowire heterostructures by Au-assisted chemical beam epitaxy.
  \emph{Nanotechnology} \textbf{2019}, \emph{30}, 094003\relax
\mciteBstWouldAddEndPuncttrue
\mciteSetBstMidEndSepPunct{\mcitedefaultmidpunct}
{\mcitedefaultendpunct}{\mcitedefaultseppunct}\relax
\EndOfBibitem
\bibitem[Svilans \latin{et~al.}(2016)Svilans, Leijnse, and
  Linke]{Svilans_thermoel_QD_1}
Svilans,~A.; Leijnse,~M.; Linke,~H. Experiments on the thermoelectric
  properties of quantum dots. \emph{Comptes Rendus Physique} \textbf{2016},
  \emph{17}\relax
\mciteBstWouldAddEndPuncttrue
\mciteSetBstMidEndSepPunct{\mcitedefaultmidpunct}
{\mcitedefaultendpunct}{\mcitedefaultseppunct}\relax
\EndOfBibitem
\bibitem[Svilans \latin{et~al.}(2015)Svilans, Burke, Svensson, Leijnse, and
  Linke]{Svilans_thermoel_QD_2}
Svilans,~A.; Burke,~A.; Svensson,~S.; Leijnse,~M.; Linke,~H. Nonlinear
  thermoelectric response due to energy-dependent transport properties of a
  quantum dot. \emph{Physica E: Low-dimensional Systems and Nanostructures}
  \textbf{2015}, \emph{82}\relax
\mciteBstWouldAddEndPuncttrue
\mciteSetBstMidEndSepPunct{\mcitedefaultmidpunct}
{\mcitedefaultendpunct}{\mcitedefaultseppunct}\relax
\EndOfBibitem
\bibitem[Staring \latin{et~al.}(1993)Staring, Molenkamp, Alphenaar, van Houten,
  Buyk, Mabesoone, Beenakker, and Foxon]{StaringEPL1993}
Staring,~A. A.~M.; Molenkamp,~L.~W.; Alphenaar,~B.~W.; van Houten,~H.; Buyk,~O.
  J.~A.; Mabesoone,~M. A.~A.; Beenakker,~C. W.~J.; Foxon,~C.~T.
  Coulomb-Blockade Oscillations in the Thermopower of a Quantum Dot.
  \emph{Europhysics Letters} \textbf{1993}, \emph{22}, 57--62\relax
\mciteBstWouldAddEndPuncttrue
\mciteSetBstMidEndSepPunct{\mcitedefaultmidpunct}
{\mcitedefaultendpunct}{\mcitedefaultseppunct}\relax
\EndOfBibitem
\bibitem[Dzurak \latin{et~al.}(1997)Dzurak, Smith, Barnes, Pepper,
  Mart\'{\i}n-Moreno, Liang, Ritchie, and Jones]{DzurakPRB1997}
Dzurak,~A.~S.; Smith,~C.~G.; Barnes,~C. H.~W.; Pepper,~M.;
  Mart\'{\i}n-Moreno,~L.; Liang,~C.~T.; Ritchie,~D.~A.; Jones,~G. A.~C.
  Thermoelectric signature of the excitation spectrum of a quantum dot.
  \emph{Phys. Rev. B} \textbf{1997}, \emph{55}, R10197--R10200\relax
\mciteBstWouldAddEndPuncttrue
\mciteSetBstMidEndSepPunct{\mcitedefaultmidpunct}
{\mcitedefaultendpunct}{\mcitedefaultseppunct}\relax
\EndOfBibitem
\bibitem[Scheibner \latin{et~al.}(2007)Scheibner, Novik, Borzenko, K\"onig,
  Reuter, Wieck, Buhmann, and Molenkamp]{ScheibnerPRB2007}
Scheibner,~R.; Novik,~E.~G.; Borzenko,~T.; K\"onig,~M.; Reuter,~D.;
  Wieck,~A.~D.; Buhmann,~H.; Molenkamp,~L.~W. Sequential and cotunneling
  behavior in the temperature-dependent thermopower of few-electron quantum
  dots. \emph{Phys. Rev. B} \textbf{2007}, \emph{75}, 041301\relax
\mciteBstWouldAddEndPuncttrue
\mciteSetBstMidEndSepPunct{\mcitedefaultmidpunct}
{\mcitedefaultendpunct}{\mcitedefaultseppunct}\relax
\EndOfBibitem
\bibitem[Prete \latin{et~al.}(2019)Prete, Erdman, Demontis, Zannier, Ercolani,
  Sorba, Beltram, Rossella, Taddei, and Roddaro]{PreteNanoLett2019}
Prete,~D.; Erdman,~P.~A.; Demontis,~V.; Zannier,~V.; Ercolani,~D.; Sorba,~L.;
  Beltram,~F.; Rossella,~F.; Taddei,~F.; Roddaro,~S. Thermoelectric Conversion
  at 30 K in InAs/InP Nanowire Quantum Dots. \emph{Nano Letters} \textbf{2019},
  \emph{19}, 3033--3039, PMID: 30935206\relax
\mciteBstWouldAddEndPuncttrue
\mciteSetBstMidEndSepPunct{\mcitedefaultmidpunct}
{\mcitedefaultendpunct}{\mcitedefaultseppunct}\relax
\EndOfBibitem
\bibitem[Thierschmann \latin{et~al.}(2013)Thierschmann, Henke, Knorr, Maier,
  Heyn, Hansen, Buhmann, and Molenkamp]{Thierschmann_thermoel_DQD}
Thierschmann,~H.; Henke,~M.; Knorr,~J.; Maier,~L.; Heyn,~C.; Hansen,~W.;
  Buhmann,~H.; Molenkamp,~L.~W. Diffusion thermopower of a serial double
  quantum dot. \emph{New Journal of Physics} \textbf{2013}, \emph{15},
  123010\relax
\mciteBstWouldAddEndPuncttrue
\mciteSetBstMidEndSepPunct{\mcitedefaultmidpunct}
{\mcitedefaultendpunct}{\mcitedefaultseppunct}\relax
\EndOfBibitem
\bibitem[Goldozian \latin{et~al.}(2019)Goldozian, Kir{\v{s}}anskas, Damtie, and
  Wacker]{Goldozian_PAT_QmeQ_testcase}
Goldozian,~B.; Kir{\v{s}}anskas,~G.; Damtie,~F.~A.; Wacker,~A. Quantifying the
  impact of phonon scattering on electrical and thermal transport in quantum
  dots. \emph{The European Physical Journal Special Topics} \textbf{2019},
  \emph{227}, 1959--1967\relax
\mciteBstWouldAddEndPuncttrue
\mciteSetBstMidEndSepPunct{\mcitedefaultmidpunct}
{\mcitedefaultendpunct}{\mcitedefaultseppunct}\relax
\EndOfBibitem
\bibitem[Josefsson \latin{et~al.}(2019)Josefsson, Svilans, Linke, and
  Leijnse]{Josefsson_theory_fits}
Josefsson,~M.; Svilans,~A.; Linke,~H.; Leijnse,~M. Optimal power and efficiency
  of single quantum dot heat engines: Theory and experiment. \emph{Phys. Rev.
  B} \textbf{2019}, \emph{99}, 235432\relax
\mciteBstWouldAddEndPuncttrue
\mciteSetBstMidEndSepPunct{\mcitedefaultmidpunct}
{\mcitedefaultendpunct}{\mcitedefaultseppunct}\relax
\EndOfBibitem
\bibitem[Blick \latin{et~al.}(1998)Blick, Pfannkuche, Haug, Klitzing, and
  Eberl]{Blick_DQD_molecular_states}
Blick,~R.~H.; Pfannkuche,~D.; Haug,~R.~J.; Klitzing,~K.~v.; Eberl,~K. Formation
  of a Coherent Mode in a Double Quantum Dot. \emph{Phys. Rev. Lett.}
  \textbf{1998}, \emph{80}, 4032--4035\relax
\mciteBstWouldAddEndPuncttrue
\mciteSetBstMidEndSepPunct{\mcitedefaultmidpunct}
{\mcitedefaultendpunct}{\mcitedefaultseppunct}\relax
\EndOfBibitem
\bibitem[Oosterkamp \latin{et~al.}(1998)Oosterkamp, Fujisawa, van~der Wiel,
  Ishibashi, Hijman, Tarucha, and
  Kouwenhoven]{Oosterkamp_DQD_regimes_molecules}
Oosterkamp,~T.~H.; Fujisawa,~T.; van~der Wiel,~W.~G.; Ishibashi,~K.;
  Hijman,~R.~V.; Tarucha,~S.; Kouwenhoven,~L.~P. Microwave spectroscopy of a
  quantum-dot molecule. \emph{Nature} \textbf{1998}, \emph{395}, 873--876\relax
\mciteBstWouldAddEndPuncttrue
\mciteSetBstMidEndSepPunct{\mcitedefaultmidpunct}
{\mcitedefaultendpunct}{\mcitedefaultseppunct}\relax
\EndOfBibitem
\bibitem[Gustavsson \latin{et~al.}(2008)Gustavsson, Studer, Leturcq, Ihn,
  Ensslin, Driscoll, and Gossard]{Gustavsson_cotunneling}
Gustavsson,~S.; Studer,~M.; Leturcq,~R.; Ihn,~T.; Ensslin,~K.; Driscoll,~D.~C.;
  Gossard,~A.~C. Detecting single-electron tunneling involving virtual
  processes in real time. \emph{Phys. Rev. B} \textbf{2008}, \emph{78},
  155309\relax
\mciteBstWouldAddEndPuncttrue
\mciteSetBstMidEndSepPunct{\mcitedefaultmidpunct}
{\mcitedefaultendpunct}{\mcitedefaultseppunct}\relax
\EndOfBibitem
\bibitem[Weinmann \latin{et~al.}(1995)Weinmann, H\"ausler, and
  Kramer]{WeinmannPRL1995}
Weinmann,~D.; H\"ausler,~W.; Kramer,~B. Spin Blockades in Linear and Nonlinear
  Transport through Quantum Dots. \emph{Phys. Rev. Lett.} \textbf{1995},
  \emph{74}, 984--987\relax
\mciteBstWouldAddEndPuncttrue
\mciteSetBstMidEndSepPunct{\mcitedefaultmidpunct}
{\mcitedefaultendpunct}{\mcitedefaultseppunct}\relax
\EndOfBibitem
\bibitem[Ono \latin{et~al.}(2002)Ono, Austing, Tokura, and
  Tarucha]{OnoScience2002}
Ono,~K.; Austing,~D.~G.; Tokura,~Y.; Tarucha,~S. Current Rectification by Pauli
  Exclusion in a Weakly Coupled Double Quantum Dot System. \emph{Science}
  \textbf{2002}, \emph{297}, 1313--1317\relax
\mciteBstWouldAddEndPuncttrue
\mciteSetBstMidEndSepPunct{\mcitedefaultmidpunct}
{\mcitedefaultendpunct}{\mcitedefaultseppunct}\relax
\EndOfBibitem
\bibitem[Johnson \latin{et~al.}(2005)Johnson, Petta, Marcus, Hanson, and
  Gossard]{JohnsonPRB2005}
Johnson,~A.~C.; Petta,~J.~R.; Marcus,~C.~M.; Hanson,~M.~P.; Gossard,~A.~C.
  Singlet-triplet spin blockade and charge sensing in a few-electron double
  quantum dot. \emph{Phys. Rev. B} \textbf{2005}, \emph{72}, 165308\relax
\mciteBstWouldAddEndPuncttrue
\mciteSetBstMidEndSepPunct{\mcitedefaultmidpunct}
{\mcitedefaultendpunct}{\mcitedefaultseppunct}\relax
\EndOfBibitem
\bibitem[Nadj-Perge \latin{et~al.}(2010)Nadj-Perge, Frolov, van Tilburg, Danon,
  Nazarov, Algra, Bakkers, and Kouwenhoven]{Nadj_Perge_spin_block_lifted}
Nadj-Perge,~S.; Frolov,~S.~M.; van Tilburg,~J. W.~W.; Danon,~J.;
  Nazarov,~Y.~V.; Algra,~R.; Bakkers,~E. P. A.~M.; Kouwenhoven,~L.~P.
  Disentangling the effects of spin-orbit and hyperfine interactions on spin
  blockade. \emph{Phys. Rev. B} \textbf{2010}, \emph{81}, 201305\relax
\mciteBstWouldAddEndPuncttrue
\mciteSetBstMidEndSepPunct{\mcitedefaultmidpunct}
{\mcitedefaultendpunct}{\mcitedefaultseppunct}\relax
\EndOfBibitem
\bibitem[Kiršanskas \latin{et~al.}(2017)Kiršanskas, Pedersen, Karlstr{\"o}m,
  Leijnse, and Wacker]{QmeQ}
Kiršanskas,~G.; Pedersen,~J.; Karlstr{\"o}m,~O.; Leijnse,~M.; Wacker,~A. QmeQ
  1.0: An open-source Python package for calculations of transport through
  quantum dot devices. \emph{Computer Physics Communications} \textbf{2017},
  \emph{221}, 317--342\relax
\mciteBstWouldAddEndPuncttrue
\mciteSetBstMidEndSepPunct{\mcitedefaultmidpunct}
{\mcitedefaultendpunct}{\mcitedefaultseppunct}\relax
\EndOfBibitem
\end{mcitethebibliography}


\providecommand{\latin}[1]{#1}
\makeatletter
\providecommand{\doi}
  {\begingroup\let\do\@makeother\dospecials
  \catcode`\{=1 \catcode`\}=2 \doi@aux}
\providecommand{\doi@aux}[1]{\endgroup\texttt{#1}}
\makeatother
\providecommand*\mcitethebibliography{\thebibliography}
\csname @ifundefined\endcsname{endmcitethebibliography}
  {\let\endmcitethebibliography\endthebibliography}{}
\begin{mcitethebibliography}{9}
\providecommand*\natexlab[1]{#1}
\providecommand*\mciteSetBstSublistMode[1]{}
\providecommand*\mciteSetBstMaxWidthForm[2]{}
\providecommand*\mciteBstWouldAddEndPuncttrue
  {\def\EndOfBibitem{\unskip.}}
\providecommand*\mciteBstWouldAddEndPunctfalse
  {\let\EndOfBibitem\relax}
\providecommand*\mciteSetBstMidEndSepPunct[3]{}
\providecommand*\mciteSetBstSublistLabelBeginEnd[3]{}
\providecommand*\EndOfBibitem{}
\mciteSetBstSublistMode{f}
\mciteSetBstMaxWidthForm{subitem}{(\alph{mcitesubitemcount})}
\mciteSetBstSublistLabelBeginEnd
  {\mcitemaxwidthsubitemform\space}
  {\relax}
  {\relax}

\bibitem[Gustavsson \latin{et~al.}(2008)Gustavsson, Studer, Leturcq, Ihn,
  Ensslin, Driscoll, and Gossard]{Gustavsson_cotunneling}
Gustavsson,~S.; Studer,~M.; Leturcq,~R.; Ihn,~T.; Ensslin,~K.; Driscoll,~D.~C.;
  Gossard,~A.~C. Detecting single-electron tunneling involving virtual
  processes in real time. \emph{Phys. Rev. B} \textbf{2008}, \emph{78}\relax
\mciteBstWouldAddEndPuncttrue
\mciteSetBstMidEndSepPunct{\mcitedefaultmidpunct}
{\mcitedefaultendpunct}{\mcitedefaultseppunct}\relax
\EndOfBibitem
\bibitem[Josefsson \latin{et~al.}(2018)Josefsson, Svilans, Burke, Hoffmann,
  Fahlvik, Thelander, Leijnse, and Linke]{QD_heat_engine_fits}
Josefsson,~M.; Svilans,~A.; Burke,~A.; Hoffmann,~E.; Fahlvik,~S.;
  Thelander,~C.; Leijnse,~M.; Linke,~H. A quantum-dot heat engine operating
  close to the thermodynamic efficiency limits. \emph{Nature Nanotechnology}
  \textbf{2018}, \emph{13}\relax
\mciteBstWouldAddEndPuncttrue
\mciteSetBstMidEndSepPunct{\mcitedefaultmidpunct}
{\mcitedefaultendpunct}{\mcitedefaultseppunct}\relax
\EndOfBibitem
\bibitem[Josefsson \latin{et~al.}(2019)Josefsson, Svilans, Linke, and
  Leijnse]{Josefsson_theory_fits}
Josefsson,~M.; Svilans,~A.; Linke,~H.; Leijnse,~M. Optimal power and efficiency
  of single quantum dot heat engines: Theory and experiment. \emph{Phys. Rev.
  B} \textbf{2019}, \emph{99}\relax
\mciteBstWouldAddEndPuncttrue
\mciteSetBstMidEndSepPunct{\mcitedefaultmidpunct}
{\mcitedefaultendpunct}{\mcitedefaultseppunct}\relax
\EndOfBibitem
\bibitem[Esposito \latin{et~al.}(2009)Esposito, Lindenberg, and den
  Broeck]{EspositoEPL2009}
Esposito,~M.; Lindenberg,~K.; den Broeck,~C.~V. Thermoelectric efficiency at
  maximum power in a quantum dot. \emph{Europhysics Letters} \textbf{2009},
  \emph{85}\relax
\mciteBstWouldAddEndPuncttrue
\mciteSetBstMidEndSepPunct{\mcitedefaultmidpunct}
{\mcitedefaultendpunct}{\mcitedefaultseppunct}\relax
\EndOfBibitem
\bibitem[Bonet \latin{et~al.}(2002)Bonet, Deshmukh, and
  Ralph]{Bonet_rate_equations}
Bonet,~E.; Deshmukh,~M.~M.; Ralph,~D.~C. Solving rate equations for electron
  tunneling via discrete quantum states. \emph{Phys. Rev. B} \textbf{2002},
  \emph{65}\relax
\mciteBstWouldAddEndPuncttrue
\mciteSetBstMidEndSepPunct{\mcitedefaultmidpunct}
{\mcitedefaultendpunct}{\mcitedefaultseppunct}\relax
\EndOfBibitem
\bibitem[Kiršanskas \latin{et~al.}(2017)Kiršanskas, Pedersen, Karlstr{\"o}m,
  Leijnse, and Wacker]{QmeQ}
Kiršanskas,~G.; Pedersen,~J.; Karlstr{\"o}m,~O.; Leijnse,~M.; Wacker,~A. QmeQ
  1.0: An open-source Python package for calculations of transport through
  quantum dot devices. \emph{Computer Physics Communications} \textbf{2017},
  \emph{221}\relax
\mciteBstWouldAddEndPuncttrue
\mciteSetBstMidEndSepPunct{\mcitedefaultmidpunct}
{\mcitedefaultendpunct}{\mcitedefaultseppunct}\relax
\EndOfBibitem
\bibitem[Goldozian \latin{et~al.}(2019)Goldozian, Kir{\v{s}}anskas, Damtie, and
  Wacker]{Goldozian_PAT_QmeQ_testcase}
Goldozian,~B.; Kir{\v{s}}anskas,~G.; Damtie,~F.~A.; Wacker,~A. Quantifying the
  impact of phonon scattering on electrical and thermal transport in quantum
  dots. \emph{The European Physical Journal Special Topics} \textbf{2019},
  \emph{227}\relax
\mciteBstWouldAddEndPuncttrue
\mciteSetBstMidEndSepPunct{\mcitedefaultmidpunct}
{\mcitedefaultendpunct}{\mcitedefaultseppunct}\relax
\EndOfBibitem
\bibitem[Wacker(2002)]{WackerPhysRep2002}
Wacker,~A. Semiconductor superlattices: a model system for nonlinear transport.
  \emph{Phys. Rep.} \textbf{2002}, \emph{357}\relax
\mciteBstWouldAddEndPuncttrue
\mciteSetBstMidEndSepPunct{\mcitedefaultmidpunct}
{\mcitedefaultendpunct}{\mcitedefaultseppunct}\relax
\EndOfBibitem
\end{mcitethebibliography}

\end{document}


\maketitle

Further information is provided here on device fabrication and measurement details including the bias offset characterization, additional experimental characterization of heated measurements in the intermediate- and weak- interdot coupling regimes, contact electron temperature estimations and modelling of the system. Finally, based on the model of the system we demonstrate disentangling of the thermoelectric effect and phonon-assisted transport.

\section{Device fabrication and measurement details\label{sec:device&meas}}

To fabricate the device shown in the main text in fig. 1(a), a nanowire (with a nominal diameter of $50\,\si{nm}$ with nominally $6\,\si{nm}$ wide InP barriers spaced at a distance of approximately $35\,\si{nm}$) was dry transferred on a degenerately doped silicon substrate with an insulating SiO$_2$ top layer. A voltage $V_{\textrm{BG}}$ can be applied to the silicon substrate for it to act as a global backgate to the nanowire. A single electron beam lithography step in combination with Ni/Au thermal evaporation is used to contact the nanowire ($S$, $D$ contacts) and define three metallic sidegates, designed as plunger gates to capacitively control the left ($V_{\textrm{L}}$) and right ($V_{\textrm{R}}$) QD. We find the top two gate arms to be electrically shorted and treat them as single gate. In addition, a sideheater electrode ($V_{\textrm{H1}}$, $V_{\textrm{H2}}$) is aligned with the DQD, which can be used as supplemental gate to electrostatically control the nanowire segment around the DQD. To heat the DQD, a voltage difference $dV_{\textrm{H}}=|V_{\textrm{H1}}-V_{\textrm{H2}}|$ is applied between the heater ends, resulting in current flow through the electrode. The highest resistive electrode segment (thinnest part) is located in close vicinity to the DQD, resulting in localized joule heating. 

All measurements are carried out in a dilution refrigerator at a base electron temperature of $\sim100\,\si{mK}$ and a measurement line resistance of $3.26\,\si{k\Omega}$. Yokogawa 7651 DC voltage sources are used to apply all relevant voltages to the device. Currents are measured on the drain contact ($D$) through a Femto DLPCA-200 I/V converter at a $1\,\si{nA/V}$ gain connected to a HP 34401A multimeter. Finally, for all measurements, a constant current offset originating from the current amplifier is determined and subtracted from the data.

\section{Device and bias offset characterization\label{A:bias_offset}}

To characterize basic transport properties of the device finite bias spectroscopy was performed by mapping the current through the DQD as a function of the plunger gate voltages $V_{\textrm{L}}$ and $V_{\textrm{R}}$ with a constant bias $V_{\textrm{SD}}$ on the source contact. Initially, for $V_{\textrm{H1}}=V_{\textrm{H2}}=V_{\textrm{BG}}=0\,\si{V}$, we found the device in a strong interdot coupling regime, suggesting low tunnel barriers. By applying a negative voltage $V_{\textrm{H1}}=V_{\textrm{H2}}=-4\,\si{V}$ to the sideheater electrode, the nanowire was locally depleted around the DQD, effectively bringing the barriers up and allowing to study more weakly coupled QDs.

\begin{figure}
\includegraphics[]{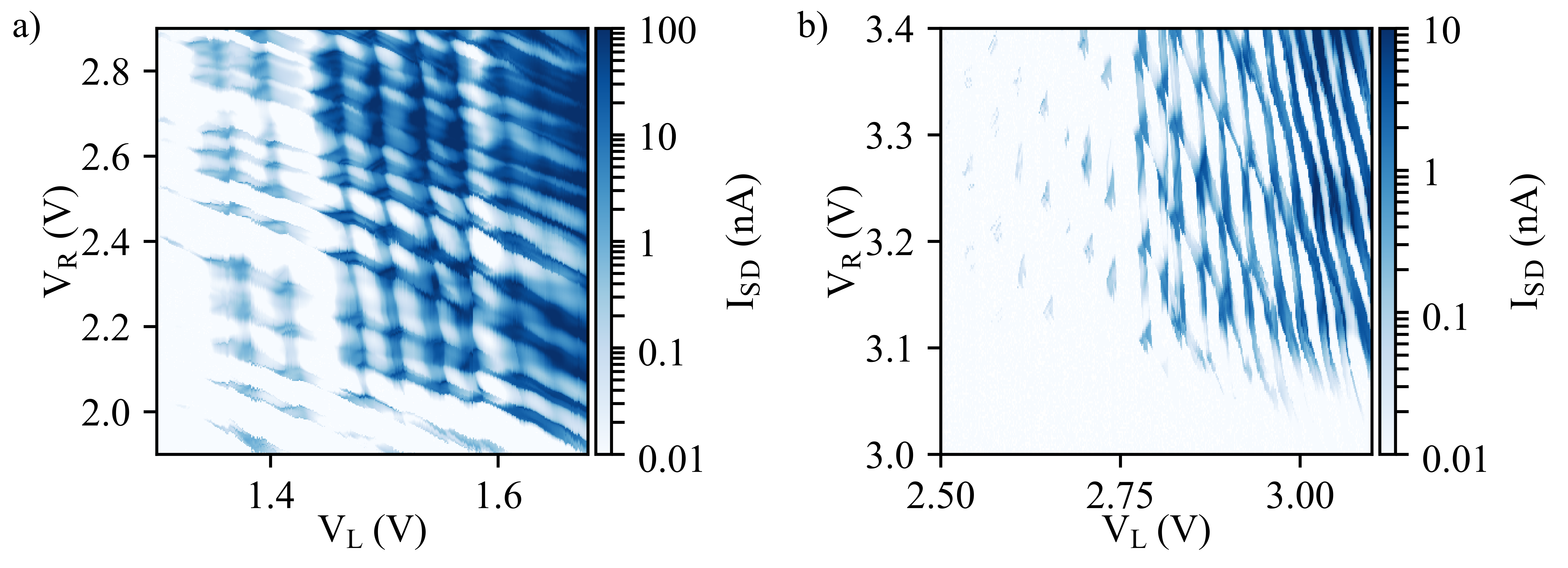}
\caption{a) Charge stability diagram in the intermediate interdot coupling regime measured with $V_{\textrm{SD}}=3\,\si{mV}$. b) Charge stability diagram in the weak interdot coupling regime measured with $V_{\textrm{SD}}=3\,\si{mV}$. The current in (a/b) is plotted on a logarithmic scale with a linear range between $\pm10\,\si{pA}.$\label{fig_S0}}
\end{figure}

Figure \ref{fig_S0}(a) shows the results of finite bias spectroscopy with $V_{\textrm{H1}}=V_{\textrm{H2}}=-4\,\si{V}$ and $V_{\textrm{BG}}=0\,\si{V}$ at $V_{\textrm{SD}}=3\,\si{mV}$. Regions of finite current are arranged in a honeycomb pattern characteristic for transport through DQDs. Charging- and charge transfer-lines are clearly visible and finite bias triangles are observed in the corners of each honeycomb cell, which indicates an intermediate interdot coupling regime. For lower plunger gate voltages and thus lower DQD occupancies more vertical charging lines are less pronounced than their more horizontal counterparts. We attribute that to an asymmetry in tunnel couplings $\Gamma_{\textrm{L}}$ between the left contact and left QD and $\Gamma_{\textrm{R}}$ between the right contact and right QD, $\Gamma_{\textrm{L}} > \Gamma_{\textrm{R}}$. For the plunger gate range studied in the main text we find lever arms of $\alpha_{\textrm{L}}=0.16\pm0.01\,\si{eV/V}$ for the left and $\alpha_{\textrm{R}}=0.06\pm0.01\,\si{eV/V}$ for the right plunger gate as well as addition energies $E_{\textrm{add,L}}=5.6\pm0.1\,\si{meV}$ and $E_{\textrm{add,R}}=2.6\pm0.1\,\si{meV}$ for the left and right QD (extracted from the bottom right honeycomb cell in fig. 2(a) in the main text).

The addition of a negative bias on the back gate was used to effectively shift the device into a weak interdot coupling regime. By applying a negative backgate we globally deplete the nanowire further and the effective tunnel barrier height is increased. Consequently, the tunnel couplings from the contacts to the QDs as well as the interdot tunnel coupling were reduced. Figure \ref{fig_S0}(b) shows the resulting charge stability diagram measured with $V_{\textrm{BG}}=-1\,\si{V}$, $V_{\textrm{H1}}=V_{\textrm{H2}}=-4\,\si{V}$ and $V_{\textrm{SD}}=3\,\si{mV}$. The absence of current along the charging lines and clear finite bias triangles forming at each triple point for $V_{\textrm{L}}<2.75\,\si{V}$ is a clear indication of a DQD in a weak interdot coupling regime. For $V_{\textrm{L}}>2.75\,\si{V}$ the device is transitioning to an intermediate and even strong interdot coupling regime as a result of effectively lower tunnel barriers for higher QD occupancies. In the weak interdot coupling regime and the plunger gate range shown in the main text in fig. 2(b), we find lever arms $\alpha_{\textrm{L}}=0.23\pm0.01\,\si{eV/V}$, $\alpha_{\textrm{R}}=0.09\pm0.01\,\si{eV/V}$ and addition energies $E_{\textrm{add,L}} =6.3\pm0.1\,\si{meV}$, $E_{\textrm{add,R}}=8.5\pm0.1\,\si{meV}$.

We note that the negative bias on the heater electrode pinches off the nanowire conductivity in close vicinity to the DQD and positive bias on the plunger gates is required to counteract this effect. As a result, occupancies cannot be estimated. Furthermore, when applying a negative backgate, both the pinch-off point and charge state of the DQD are shifted and it is not possible to measure in the same charge configuration for weak and intermediate interdot coupling. For the temperature gradient induced transport phenomena discussed in this work, only a TLS is required and the DQD occupancy can be neglected. We thus pick a stable charge configuration in the weak and intermediate interdot coupling regime (see fig. 2(a) and (b) in the main text) to serve as operation points for heated measurements.

In this study we aim at measuring pure heat driven currents at zero bias across a DQD. Consequently, a source-drain bias offset originating from the experimental setup has to be accounted for. To characterize this bias offset, we minimize the current $I_{\textrm{SD}}$ through the unheated DQD in a highly conductive state. We therefore measure $I_{\textrm{SD}}$ on a triple point (TP) and vary $V_{\textrm{SD}}$ in a narrow range around $0\,\si{V}$ and find currents to vanish at $V_{\textrm{SD}}=-26\pm 2\,\si{\mu V}$ in the weak and $V_{\textrm{SD}}=-44\pm 2\,\si{\mu V}$ in the intermediate interdot coupling regime. Repetition of this calibration measurement reveals a slow drift of the bias offset over time in the order of $\pm 2\,\si{\mu V}$. It is noteworthy, that currents induced by this drift manifest in a more pronounced manner if the DQD is in a high conductive state and are mainly present in the intermediate interdot coupling regime. In contrast, in the higher resistive weak interdot coupling regime the effect is neglegible off the exact TP location. Throughout the heated experiments, we frequently repeated control measurements to exclude the possibility of larger, unnoticed drifts.

Based on the findings described in this section, we adjusted $V_{\textrm{SD}}$ by $-26\,\si{\mu V}$ in the weak and $-44\,\si{\mu V}$ in the intermediate interdot coupling regime in all experimental measurements presented in this work.

\subsection{Intermediate interdot coupling regime\label{sec:intermediate_regime}}

\begin{figure}
\includegraphics[width=8cm]{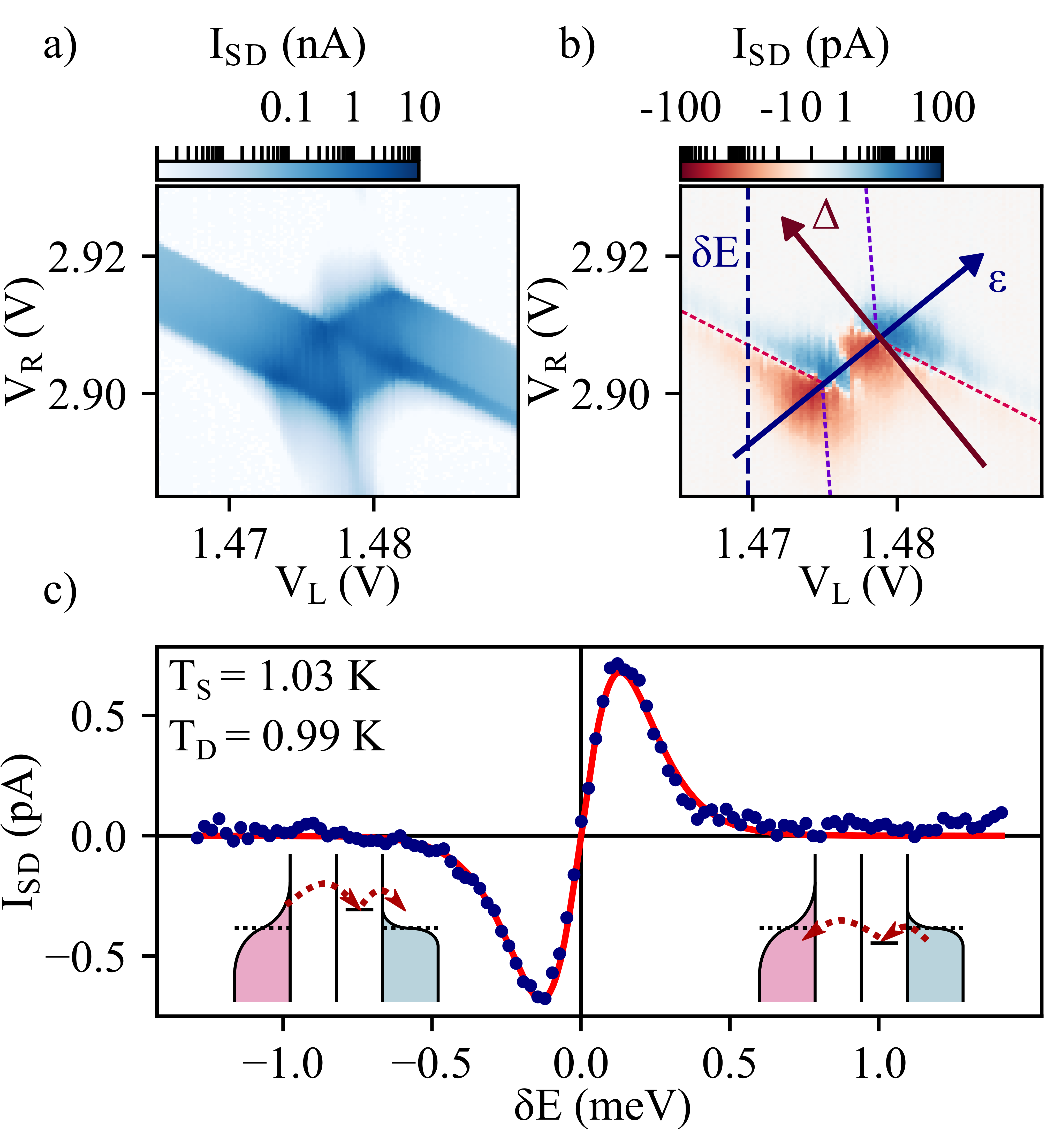}
\caption{a) Charge stability diagram around the center set of TPs in fig. 2(a) in the main text (also marked by a grey rectangle in fig. 2(c) in the main text at $V_{\textrm{SD}}=1\,\si{mV}$. b) Heated equivalent of (a) at $V_{\textrm{SD}}=0\,\si{V}$ and $dV_{\textrm{H}}=1\,\si{V}$. Characteristic energy axes $\epsilon$ and $\Delta$ are marked by blue and red arrows and dashed pink/purple lines outline regions of constant charge on the DQD. The current is plotted logarithmic with a linear range between $-1\,\si{pA}$ and $1\,\si{pA}$. c) $I_{\textrm{SD}}$ along the linecut $\delta E$, marked in (b). Blue dots are extracted from (b). The red solid line is a fit to the data to estimate contact electron temperatures $T_{\textrm{S}}=1.03\,\si{K}$ and $T_{\textrm{D}}=0.99\,\si{K}$. The transport mechanism resulting in a negative (positive) current peak along the cutline is schematically illustrated in the inset.\label{fig_S1}}
\end{figure}

In this section, we present a more detailed characterization of the heat driven transport signals obtained in the intermediate interdot coupling regime. We therefore focus on the center set of TPs in fig. 2(a) in the main text and perform high resolution scans, mapping the current $I_{\textrm{SD}}$ as a function of a narrow plugner gate range in biased and heated measurements. Figure \ref{fig_S1}(a) shows the result of a high resolution finite bias spectroscopy measurement at $V_{\textrm{SD}}=1\,\si{mV}$, showing a charge transition in the intermendiate interdot coupling regime. The asymmetry in tunnel couplings, $\Gamma_{\textrm{L}}>\Gamma_{\textrm{R}}$, leads to more pronounced horizontal charging lines compared to their vertical counterparts. Traces of an excited state are observed within the finite bias triangles and along the charging lines as regions of increased current.

The heated equivalent of fig. \ref{fig_S1}(a) at $V_{\textrm{SD}}=0\,\si{V}$ and $dV_{\textrm{H}}=1\,\si{V}$ is shown in fig. \ref{fig_S1}(b) and presents a high resolution scan of the range indicated by a grey dotted rectangle in fig. 2 (c) in the main text. By comparing the heated measurement in figure \ref{fig_S1}(b) to the bias measurement in fig. \ref{fig_S1}(a) we find currents in the same areas within the charge stability diagram. A distinct difference is the appearance of currents of opposite polarities in heated measurements, which is characteristic for temperature gradient driven effects. Pronounced current polarity reversal is observed upon crossing the pink dashed charging lines (travelling along $\delta E$) and parallel to the energy axis $\epsilon$, when crossing one of the TPs. Along the level detunement axis $\Delta$ no comparably pronounced effect is observed. In agreement with the discussion in the main text, in vicinity of the TPs, this behaviour is covered by the thermoelectric effect governing transport along $\epsilon$, mixed with weak traces of PAT which induce additional current modulation along $\Delta$. 

On the pink dashed charging lines in fig. \ref{fig_S1}(b), which outline a change of charge on the right dot, currents are observed. On these lines, a level of the right QD is aligned with the electrochemical potentials of the contacts $\mu=\mu\textrm{S}=\mu_\textrm{D}$ and no level of the left QD is available close to $\mu$. As a result of the high interdot tunnel coupling $\Omega$, the DQD can be described as a single QD with an effective left tunnel coupling $\Gamma_{\textrm{L,eff}}\approx\Gamma_{\textrm{L}}\Omega^2/\Delta^2$ along the pink dashed lines \cite{Gustavsson_cotunneling}. Neglecting plunger gate to QD cross-couplings, we expect the device to behave like a thermoelectrically driven single QD along $\delta \textrm{E}=\mu-\textrm{E}_\textrm{R}$ upon crossing the charging lines, as illustrated in the insets of fig. \ref{fig_S1}(c). We note that along the charging lines, PAT conditions are not fulfilled because no pair of levels below and above $\mu$ is within the phonons' accessible energy window. Thus, the thermoelectric effect is the only available transport mechanism. Along the purple dashed charging lines in fig. \ref{fig_S1}(b) the thermoelectric effect is suppressed due to the reservoir to QD tunnel coupling asymmetry.

Blue circles in fig. \ref{fig_S1}(c) show extracted data points along the dotted cut-line $\delta\textrm{E}$ in fig. \ref{fig_S1}(b). By considering the detected current polarities, we conclude $T_{\textrm{L}}>T_{\textrm{R}}$. For $\delta\textrm{E}<0$ ($\delta\textrm{E}>0$), the level $\textrm{E}_{\textrm{R}}$ is above (below) $\mu$ (left/right inset in fig. \ref{fig_S1}(c)) and the difference in occupation in the contacts due to $T_{\textrm{S}}>T_{\textrm{D}}$ results in electron flow from hot to cold (cold to hot). In recent work by Josefsson et al. \cite{QD_heat_engine_fits, Josefsson_theory_fits} on thermoelectric transport across single QDs comparable current traces are fitted to extract contact electron temperatures. We use a simplified approximation and compare our data to the calculated thermoelectric current through a spinless single QD \cite{EspositoEPL2009} (solid red line, fig. \ref{fig_S1}(c)) and obtain a temperature estimate of $T_{\textrm{S}}=1.03\,\si{K}$ and $T_{\textrm{D}}=0.99\,\si{K}$ for the depicted cutline. Details of the contact electron temperature estimation are given later in this document.

\subsection{Weak interdot coupling regime\label{sec:weak_regime}}

\begin{figure}
\includegraphics[width=8cm]{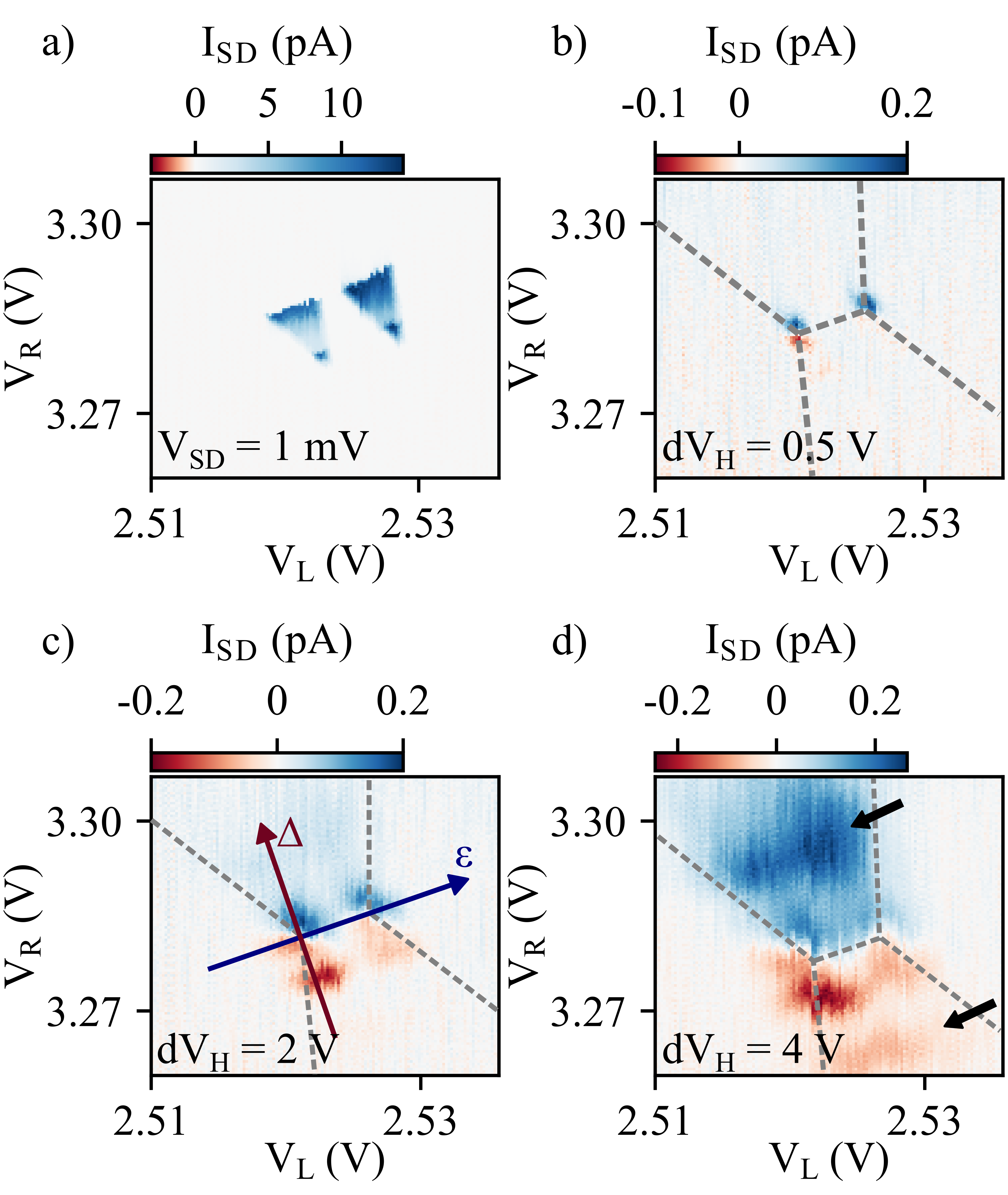}
\caption{a) High-resolution finite bias spectroscopy at $V_{\textrm{SD}}=1\,\si{mV}$ in the weak interdot tunnel coupling regime on the top left set of TPs in fig. 2(b) in the main text. b/c/d) Heated measurements  at $V_{\textrm{SD}}=0\,\si{V}$ and b) $dV_{\textrm{H}}=0.5\,\si{V}$, c) $dV_{\textrm{H}}=2\,\si{V}$, d) $dV_{\textrm{H}}=4\,\si{V}$ in the same plunger gate range as shown in (a). Each datapoint is the result of six-time averaging. The honeycomb cell of the charge stability diagram is outlined by dashed lines as guide to the eye. Characteristic energy axes $\epsilon$ and $\Delta$ are labelled by a blue and red arrow respectively in (c). Black arrows in (d) indicate the presence of resonances within the heat driven current signal for high $dV_{\textrm{H}}$.\label{fig_S2}}
\end{figure}

In this section, we characterize the heat driven current signals obtained in the weak interdot coupling regime. Here, we focus on the top left set of TPs in fig. 2(b) in the main text and study high-resolution scans of the charge stability diagram. The chosen plunger gate range corresponds to the range outlined by a dotted grey rectangle in fig. 2(d) in the main text. Figure \ref{fig_S2}(a) shows the result of finite bias spectroscopy at $V_{\textrm{SD}}=1\,\si{mV}$. Current is only detected within triangular shaped regions extending out of each TP which is characteristic for a weakly coupled DQD and traces of an excited state at $0.8\,\si{meV}$ are observed. Figure \ref{fig_S2}(b/c/d) show the purely heat driven counterpart of (a) for different heating bias $dV_\textrm{H}$.

First, we study the characteristic signal shape for the weak interdot coupling regime observed for moderate heating, $dV_{\textrm{H}}=2\,\si{V}$, as shown in fig. \ref{fig_S2}(c). A higher $dV_{\textrm{H}}$ compared to the intermediate interdot coupling regime is chosen to obtain sufficient phonon energies to study PAT at large $\Delta$, where thermoelectric currents are suppressed. Within the observed signal a clear reversal of current polarity is observed perpendicular to the charge transfer line, along $\Delta$. For positive and negative $\Delta$, current readings imply electron transport against the level detunement, from energetically lower to higher DQD levels. This inelastic electron transport can therefore not be caused by a thermoelectric effect or phonon emission. Further ruling out cotunneling due to the small $\Omega$ in the weak interdot coupling regime, this suggests that PAT by absorbing a phonon supplied by the heater electrode is the predominant transport mechanism. The relevant PAT process is discussed in the main text in fig. 1(d). In agreement with the discussions in the main text, we find current contributions attributed to PAT-processes to be essentially limited by the boundaries of regions of constant charge (grey dashed lines) in the charge stability diagram.

Clear indications of the thermoelectric effect are only present for small $\Delta\approx 0$. Finite currents extend noticeably beyond the boundaries of regions of constant charge and we find negative currents for $\epsilon < 0$ and positive currents for $\epsilon > 0$. In addition, when continuing to travel along $\epsilon$ for small $\Delta\approx 0$, a current polarity reversal parallel to the axis $\epsilon$ upon crossing each TP is observed.

In combination, the measured current signal in fig. \ref{fig_S2}(c) captures closely the expectations for a weakly coupled DQD in the presence of a temperature gradient $T_{\textrm{ph}}>T_{\textrm{S}}>T_{\textrm{D}}$. For large $\Delta$ PAT dominates transport and pure thermoelectric signals are found outside the boundaries of PAT and for small $\Delta$, along $\epsilon$. In close vicinity to the TPs, within a region of constant charge on the DQD, the detected currents are the result of a superposition of both thermoelectric and PAT contributions. We stress that in contrast to the intermediate interdot coupling regime, PAT is the predominant driving mechanism in the weak interdot coupling regime. To demonstrate that difference further, we repeat the measurement of fig. \ref{fig_S2}(c) at a reduced $dV_{\textrm{H}}=0.5\,\si{V}$, which is shown in fig. \ref{fig_S2}(b). While in the intermediate interdot coupling regime for $dV_{\textrm{H}}=1\,\si{V}$ (fig. \ref{fig_S1}(b)) transport is almost exclusively mediated by the thermoelectric effect, we already find the most pronounced current polarity reversal along $\Delta$ in fig. \ref{fig_S2}(b), underlining the increased relevance of PAT for lower $\Omega$. 

Finally, we increase the heating bias further to $dV_{\textrm{H}}=4\,\si{V}$ as shown in fig. \ref{fig_S2}(d). We find the signal for high heating bias to be conceptually comparable to the signal for moderate in fig. \ref{fig_S2}(c) with the exception of pronounced resonances emerging within the PAT current signal (labelled by black arrows). These features are discussed in detail in the main text.

\section{Contact electron temperature estimation\label{A:T_estimate}}

The contact electron temperatures $T_{\textrm{S}}$ and $T_{\textrm{D}}$ are important parameters that need to be extracted from the experiment in order to obtain realistic simulations of our device. Heated measurements in our system depend on the three temperatures $T_{\textrm{S}}$, $T_{\textrm{D}}$ and $T_{\textrm{ph}}$. To reduce the amount of free parameters for fits, we aim at obtaining contact electron temperatures from current signals along charging lines in heated measurements in the intermediate interdot coupling regime. In line with the discussions in the main text, heat driven currents along the charging lines result from the thermoelectric effect only and therefore do not depend on $T_{\textrm{ph}}$.

\begin{figure}
\includegraphics[width=\textwidth]{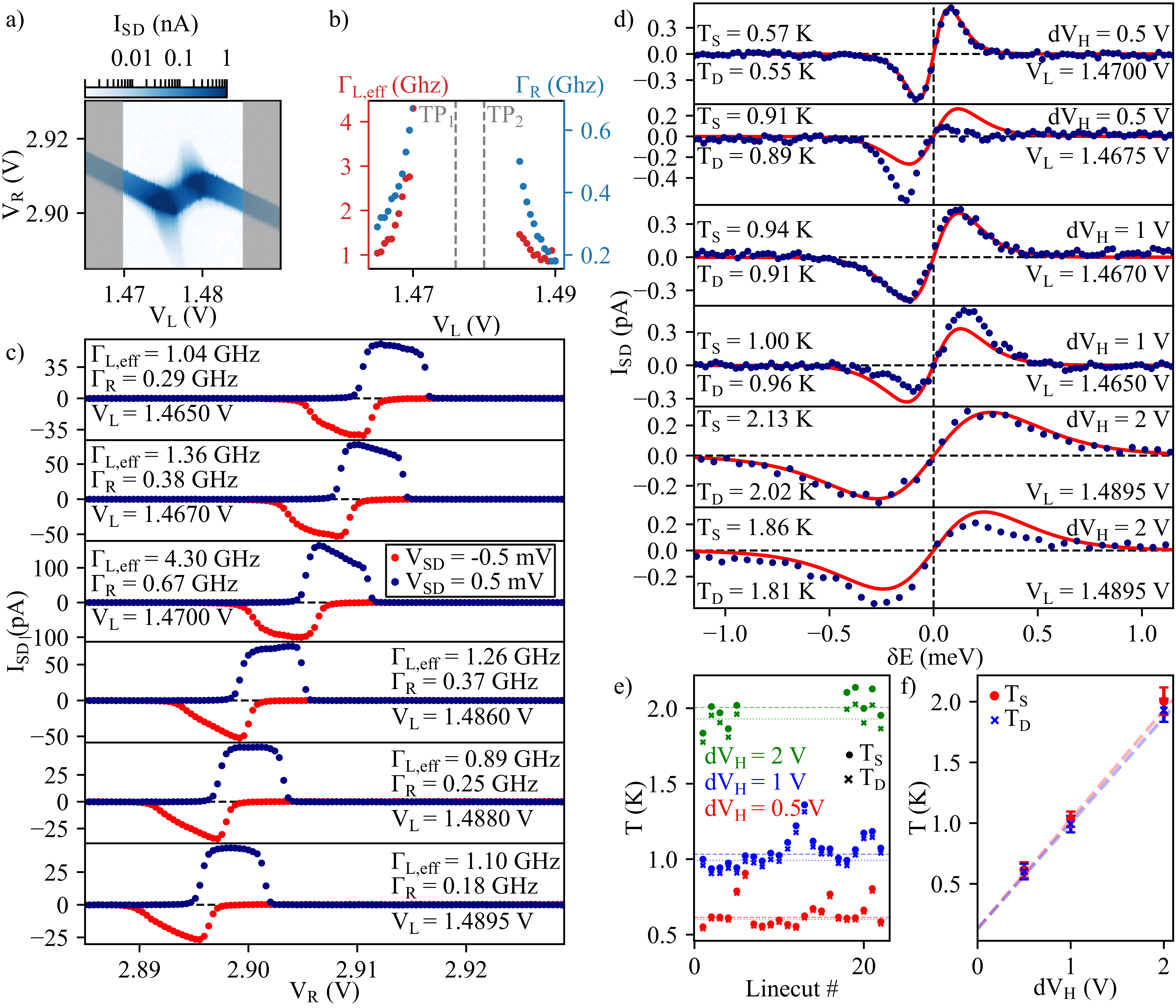}
\caption{a) Measured charge stability diagram of the same range discussed in fig. \ref{fig_S1}. Cutlines to estimate the tunnel couplings and contact electron temperatures are taken in the grey shaded areas. b) Tunnel couplings $\Gamma_{\textrm{L,eff}}$ and $\Gamma_{\textrm{R}}$ along the two pronounced charging lines in (a) as a function of $V_{\textrm{L}}$. c) Selection of data along $V_{\textrm{R}}$ at constant $V_{\textrm{L}}$ for positive and negative bias. Tunnel couplings are extracted from saturation currents. d) Selection of zero-bias linecuts at different $dV_{\textrm{H}}$. Dots mark data points and solid red lines represent fits based on rate equations for a spinless single QD to extract $T_{\textrm{S}}$ and $T_{\textrm{D}}$. e) Extracted $T_{\textrm{S}}$ and $T_{\textrm{D}}$ values from fits at different $dV_{\textrm{H}}$. Dashed and dotted lines represent the median values for $T_{\textrm{S}}$ and $T_{\textrm{D}}$ at different $dV_{\textrm{H}}$ respectively. f) Median values for $T_{\textrm{S}}$ and $T_{\textrm{D}}$ plotted against $dV_{\textrm{H}}$. The dashed red and blue lines represent our estimate for contact temperatures $T_{\textrm{S}}\approx\ 0.9\,\si{K/V}\cdot dV_{\textrm{H}}+130\,\si{mK}$ and $T_{\textrm{D}}\approx\ 0.87\,\si{K/V}\cdot dV_{\textrm{H}}+130\,\si{mK}$. Error bars represent the median absolute deviation.\label{fig_S3}}
\end{figure}

Figure \ref{fig_S3}(a) shows the measured charge stability diagram at $dV_{\textrm{H}}=0\,\si{V}$ and $V_{\textrm{SD}}=0.5\,\si{mV}$ in the same range as the heated measurement in fig. \ref{fig_S1}. As a result of the asymmetry $\Gamma_{\textrm{L}}>\Gamma_{\textrm{R}}$, only the charging lines where upon crossing the occupancy of the right QD varies are clearly visible. To estimate contact electron temperatures at different $dV_{\textrm{H}}$, we keep a distance to the TPs to avoid superposition of different effects and focus on the visible charging lines only. The ranges of the charge stability diagram used for the temperature estimation are marked by grey shaded regions in fig. \ref{fig_S3}(a).

As described in the main text, if the plunger-gate to QD cross-couplings are neglected, our DQD acts as a single QD with the (effective) tunnel couplings $\Gamma_{\textrm{L,eff}}$ and $\Gamma_{\textrm{R}}$ when travelling across the charging line along $V_{\textrm{R}}$ at a constant $V_{\textrm{L}}$. In heated measurements at no driving bias current readings along such cut-lines recover the thermoelectric signal of effectively a single QD with a current polarity reversal when the QD level is energetically pulled from above to below $\mu$ as demonstrated in fig. \ref{fig_S1}(c). In recent work on single QDs \cite{Josefsson_theory_fits,QD_heat_engine_fits}, fits to comparable cut-lines have been used to obtain the contact electron temperatures $T_{\textrm{S}}$ and $T_{\textrm{D}}$. Within the grey shaded regions, we extract current readings along 22 cut-lines (11 at $V_{\textrm{L}}<1.47\,\si{V}$ and 11 at $V_{\textrm{L}}>1.485\,\si{V}$) from heated measurements at no bias for $dV_{\textrm{H}}=0.5\,\si{V}$ and $dV_{\textrm{H}}=1\,\si{V}$. We further extract data along 10 cut-lines from a measurement at $dV_{\textrm{H}}=2\,\si{V}$, where an increased distance to the TPs is required to avoid overlap of current signals along charging lines and around TPs. A selection of these cut-lines for different $V_{\textrm{L}}$ and $dV_{\textrm{H}}$ is shown in fig. \ref{fig_S3}(d).

In line with the discussions by Josefsson et al. \cite{Josefsson_theory_fits}, depending on the QD occupancy, either the ingoing or outgoing tunneling process limits the current in heated measurements and as a result an asymmetry between the positive and negative signal amplitude is expected. Within our cut-line data however, no clear trend in the asymmetry is present. We attribute this to the drift of $\pm 2\,\si{\mu V}$ over time in the offset bias. This drift is of comparable magnitude to $k_{\textrm{B}}\Delta T/e$ and therefore cloaks the expected amplitude difference. With no experimental access to exact QD occupancies, we consequently opt for a simple approximation instead of fitting the cut-line data based on a more complex model. To obtain an estimate of $T_{\textrm{S}}$ and $T_{\textrm{D}}$, we compare the experimental results to the current calculated by rate equations for a single level QD, neglecting spin. According to Esposito et al. \cite{EspositoEPL2009}, the calculated current for that simplified case is given by \begin{equation}
I_{\textrm{SD}}=-e\dfrac{\Gamma_{\textrm{L,eff}}\Gamma_{\textrm{R}}}{\Gamma_{R}+\Gamma_{\textrm{L,eff}}}\cdot(f_{\textrm{D}}-f_{\textrm{S}})
\label{EqFIT}
\end{equation} where $f_{\textrm{S/D}}=[e^{-\delta E/(k_{\textrm{B}}T_{\textrm{S/D}})}+1]^{-1}$ is the Fermi distribution.

In order to compare currents calculated by eq. \ref{EqFIT} to the heated experimental data, we first extract the (effective) tunnel couplings $\Gamma_{\textrm{L,eff}}$ and $\Gamma_{\textrm{R}}$ from the experiment. Unheated, bias driven currents through a single spin degenerate level of a QD saturate at $I_+$/$I_-$ and can be described by\begin{equation}\begin{split}
I_+ =&\, 2e\dfrac{\Gamma_{\textrm{L,eff}}\Gamma_{\textrm{R}}}{2\Gamma_{R}+\Gamma_{\textrm{L,eff}}}\\
I_- =& -2e\dfrac{\Gamma_{\textrm{L,eff}}\Gamma_{\textrm{R}}}{\Gamma_{R}+2\Gamma_{\textrm{L,eff}}}
\label{EqCouplings}
\end{split}\end{equation} for large positive/negative bias $|V_{\textrm{SD}}|\gg k_{\textrm{B}}T/e$\cite{Bonet_rate_equations}. Consequently, we measure the current along the same 22 cut-lines within the grey dashed areas at $dV_{\textrm{H}}=0\,\si{V}$ and $V_{\textrm{SD}}=\pm 0.5\,\si{mV}$ (small enough to avoid contributions of the first excited state (ES) at $0.8\,\si{meV}$). A selection of the resulting cut-line pairs at different $V_{\textrm{L}}$ is shown in fig. \ref{fig_S3}(c). We note that for $V_{SD}=0.5\,\si{mV}$ the offset bias drift is negligible and currents saturate for positive bias. The less pronounced saturation behaviour for negative bias can possibly be attributed to the non-constant density of states in the nanowire and we use the maximum value for $I_-$. With eq. \ref{EqCouplings} and based on the asymmetry between $I_+$ and $I_-$ we obtain the (effective) tunnel couplings $\Gamma_{\textrm{L,eff}}$ and $\Gamma_{\textrm{R}}$, which are plotted as a function of $V_{\textrm{L}}$ in fig. \ref{fig_S3}(b).

With the tunnel couplings from fig. \ref{fig_S3}(b) and eq. \ref{EqFIT} we then fit the unbiased, heated cut-line data to estimate $T_{\textrm{S}}$ and $T_{\textrm{D}}$. Figure \ref{fig_S3}(d) shows a selection of heated cut-line data (blue dots) and fits (solid red lines) for different $dV_{\textrm{H}}$ and $V_{\textrm{L}}$. We stress that as a direct consequence of the offset bias drift and eq. \ref{EqFIT} not capturing the full physics of the system, not all fits are in good agreement with the experimental data (see panel 2, 4 and 6, fig. \ref{fig_S3}(d)). Statistically, however, we still probe the amplitude and width of the thermoelectric signal along $\delta E$, which directly relate to the average temperature $\widetilde{T}$ and temperature difference $\Delta T = T_{\textrm{S}}-T_{\textrm{D}}$, respectively. The values for $T_{\textrm{S}}$ and $T_{\textrm{D}}$ for all fits at different $dV_{\textrm{H}}$ are plotted in fig. \ref{fig_S3}(e), demonstrating a clear increase of $\widetilde{T}$ and a small increase of $\Delta T$ with increasing $dV_{\textrm{H}}$. To minimize the impact of outliers due to bad fits, we look at median values for $T_{\textrm{S}}$ (dashed lines in fig. \ref{fig_S3}(e)) and $T_{\textrm{D}}$ (dotted lines in fig. \ref{fig_S3}(e)), which are plotted against $dV_{\textrm{H}}$ in fig. \ref{fig_S3}(f). On basis of these median temperatures, we estimate $T_{\textrm{S}}$ and $T_{\textrm{D}}$ (dashed lines in fig. \ref{fig_S3}(f)) and find
  \begin{equation}\begin{split}
    T_{\textrm{S}}\approx&\ 0.9\,\si{K/V}\cdot dV_{\textrm{H}}+130\,\si{mK}\\
    T_{\textrm{D}}\approx&\ 0.87\,\si{K/V}\cdot dV_{\textrm{H}}+130\,\si{mK}\, ,
    \label{EqTLTRfit}
\end{split}\end{equation}
where the small temperature difference $\Delta T$ is in good agreement with our expectations for a nearly symmetric heating effect of the heater electrode.

\section{Simulations \label{sec:modelling}}

For modelling of the DQD system, we use the simulation tool QMEQ [for \textit{Quantum Master Equation for Quantum dot transport calculations}] \cite{QmeQ} with the recent inclusion of phonon scattering \cite{Goldozian_PAT_QmeQ_testcase}. All results shown are calculated with the Pauli master equation, after checking  at selected points, that the inclusion of coherences by the Redfield approach hardly changed the results. A summary of the parameters used is given in table \ref{TableSimParameters}.

\subsection{Hamiltonian and input parameters}
For the left dot we consider 2 spin degenerate levels (g,e) with ground level energy $E_L$ and the excited level energy $E_\textrm{L}+\Delta E_\textrm{L}$ (the right dot is treated analogously with $L\to R$). With a constant intradot Coulomb interaction $U$ and corresponding exchange term $U_\mathrm{ex}$
we have the Hamiltonian in occupation number representation
\[\begin{split}
\hat{H}_\textrm{R}=& \sum_\sigma E_\textrm{L} \hat{a}^\dag_{Lg\sigma}\hat{a}_{Lg\sigma}
+(E_\textrm{L}+\Delta E_\textrm{L}) \hat{a}^\dag_{Le\sigma}\hat{a}_{Le\sigma}
+\frac{U}{2}\sum_{n\sigma n'\sigma'} \hat{a}^\dag_{Ln\sigma}\hat{a}^\dag_{Ln'\sigma'}\hat{a}_{Ln'\sigma'}\hat{a}_{Ln\sigma}\\
&+\frac{U_\mathrm{ex}}{2}\sum_{n\sigma n'\sigma'} \hat{a}^\dag_{Ln\sigma}\hat{a}^\dag_{Ln'\sigma'}\hat{a}_{L\bar{n}'\sigma'}\hat{a}_{L\bar{n}\sigma}
\end{split}\]
where $\sigma$ denotes the spin, $n=e,g$,  and  
$\bar{n}$ is defined by $\bar{g}=e,\bar{e}=g$. 

The different dots are coupled by 
\[ \hat{H}_\textrm{coup}=
\sum_{nn',\sigma} \Omega \hat{a}^\dag_{Rn'\sigma}\hat{a}_{Ln\sigma}+h.c.
+\sum_{nn',\sigma\sigma'} U_n \hat{a}^\dag_{Ln\sigma}\hat{a}^\dag_{Rn'\sigma'}\hat{a}_{Rn'\sigma'}\hat{a}_{Ln\sigma}
\]
with the tunnel coupling $\Omega$ and the Coulomb interaction $U_n$ between neighboring dots.
We neglected the (tiny) exchange and further terms for different dots.

The simulations are based on a master equation in the basis of eigenstates for $\hat{H}_L+\hat{H}_R+\hat{H}_\textrm{coup}$. The coupling to contacts and phonons is first evaluated in the local basis and then transformed to these states which is automatically done within the QMEQ package \cite{QmeQ}.

The values $\Delta E_{\textrm{L/R}}$, $U$, $U_\mathrm{ex}$, and 
$U_n$ are fitted to different features in the charging diagrams like fig. 2(b) of the article and given in Table \ref{TableSimParameters}. We note that $U$ or $U_n$ can be expressed by the charging energy $e^2/(4\pi\epsilon r)$ with  $r=12\, \si{\nano\m}$ or $68\, \si{\nano\m}$, which are typical intra/inter-dot distances for our system, respectively.

\begin{table}[t]
\begin{tabular}{|l|l|}
\hline
$\Delta E_\textrm{L}=1.7$ meV &$\Delta E_\textrm{R}=0.8$ meV\\
$U=8$ meV &$U_\mathrm{ex}=0.4$ meV\\
$U_n=1.4\, \si{\milli\eV}$ &$\Omega=14\, \si{\micro\eV}$\\
$\Gamma_\textrm{L}=0.15\, \si{\micro\eV}$ &$\Gamma_\textrm{R}=0.1\,\si{\micro\eV}$\\ 
\hline
\end{tabular}
\caption{Simulation parameters used for the weak interdot coupling regime.
For intermediate interdot coupling we change $\Omega=0.17$ meV, $\Gamma_\textrm{L}=6\, \si{\micro\eV}$, $\Gamma_\textrm{R}=4\, \si{\micro\eV}$, and $U_n=0.4\, \si{\milli\eV}$. }
\label{TableSimParameters}
\end{table}

The tunnel coupling $\Omega$ is evaluated for the ground Wannier level 
with energy $E_G$ in an 40 nm InAs/5 nm InP superlattice following \cite{WackerPhysRep2002}. 
Based on Fermi's golden rule, we estimate the escape rate to the left lead as $\Gamma_\textrm{R}=\frac{2\pi}{\hbar}\Omega^2 D(E_G)$, where $D(E)$ is the one-dimensional density of states in the lead. We get $\Gamma_\textrm{R}\approx 0.1 \si{\micro\eV}$ and we use a slightly larger value for $\Gamma_\textrm{L}$ as indicated by the experiment. We note, that $\Gamma_{L/R}$ and $\Omega$ essentially quantify the magnitude of current at direct tunneling lines and the good quantitative agreement demonstrates the validity of these estimates.

For the intermediate interdot coupling regime, the dots have a higher occupation, and we use the forth Wannier state resulting in larger values of $\Omega$ and $\Gamma_{\textrm{L/R}}$.
From fig.~2(a) of the article, we also identify a smaller value of $U_n$, which 
may be related to increased screening for larger dot occupations. For simplicity, 
we refrained form a new fit of the other values.

Phonon scattering is implemented as in ref.~\cite{Goldozian_PAT_QmeQ_testcase}. However we use the phonon spectral function
  \begin{equation}
    J(E)=\frac{0.05E^3}{\si{\meV^2}}e^{-E^2a^2/15\si{\meV^2.\nm^2}}
  \end{equation}
  which is appropriate for bulk acoustic phonons. We use the Gaussian radius $a=5\,\si{\nm}$ and interdot distance $d=45\,\si{\nm}$ for the wave functions. For the phonon temperature we apply
  \begin{equation}
    T_\textrm{ph}=1.2\,\si{K/V}\cdot dV_{\textrm{H}}+130\,\si{mK}\, ,\label{EqTphfit}
    \end{equation}
assuming that the bias applied to the side-heater electrode $dV_{\textrm{H}}$ has a stronger impact on the phonons in the DQD than on the electrons in the leads, compare eq.~(\ref{EqTLTRfit}).

\subsection{Two particle states of right dot}
By diagonalisasing $\hat{H}_R$, we find the  
following two-particle states
\begin{description}
\item{ground singlet} with energy
$E_g^S=2E_\textrm{R}+\Delta E_\textrm{R}+U-\sqrt{\Delta E_\textrm{R}^2+U_\mathrm{ex}^2}$
\item{triplet} with energy
$E^T=2E_\textrm{R}+\Delta E_\textrm{R}+U -U_\mathrm{ex}$
\item{first excited singlet} with energy
$E^S_I=2E_\textrm{R}+\Delta E_\textrm{R}+U +U_\mathrm{ex}$
 \item{second excited singlet} with energy
$E_{II}^S=2E_\textrm{R}+\Delta E_\textrm{R}+U+\sqrt{\Delta E_\textrm{R}^2+U_\mathrm{ex}^2}$
\end{description}
which are used to determine the resonance conditions for tunneling
$GS(1,1)\to GS(0,2)$. Based on the equality of the ground state energies and neglecting the tunnel coupling $\Omega$ we get
\begin{equation}\begin{split}
E_\textrm{L}+E_\textrm{R}+U_n&=2E_\textrm{R}+\Delta E_\textrm{R}+U-\sqrt{\Delta E_\textrm{R}^2+U_\mathrm{ex}^2}
\Rightarrow
E_\textrm{L}=E_\textrm{R}+\Delta_0\\
&\textrm{with }
\Delta_0=U-U_n+\Delta E_\textrm{R}-\sqrt{\Delta E_\textrm{R}^2+U_\mathrm{ex}^2}
\end{split}\end{equation}
which provides $\Delta_0$ used in the main article. 
Similar calculations provide the other resonances.

\section{Disentangling the TE effect and PAT}

\begin{figure}
\includegraphics[width=\textwidth]{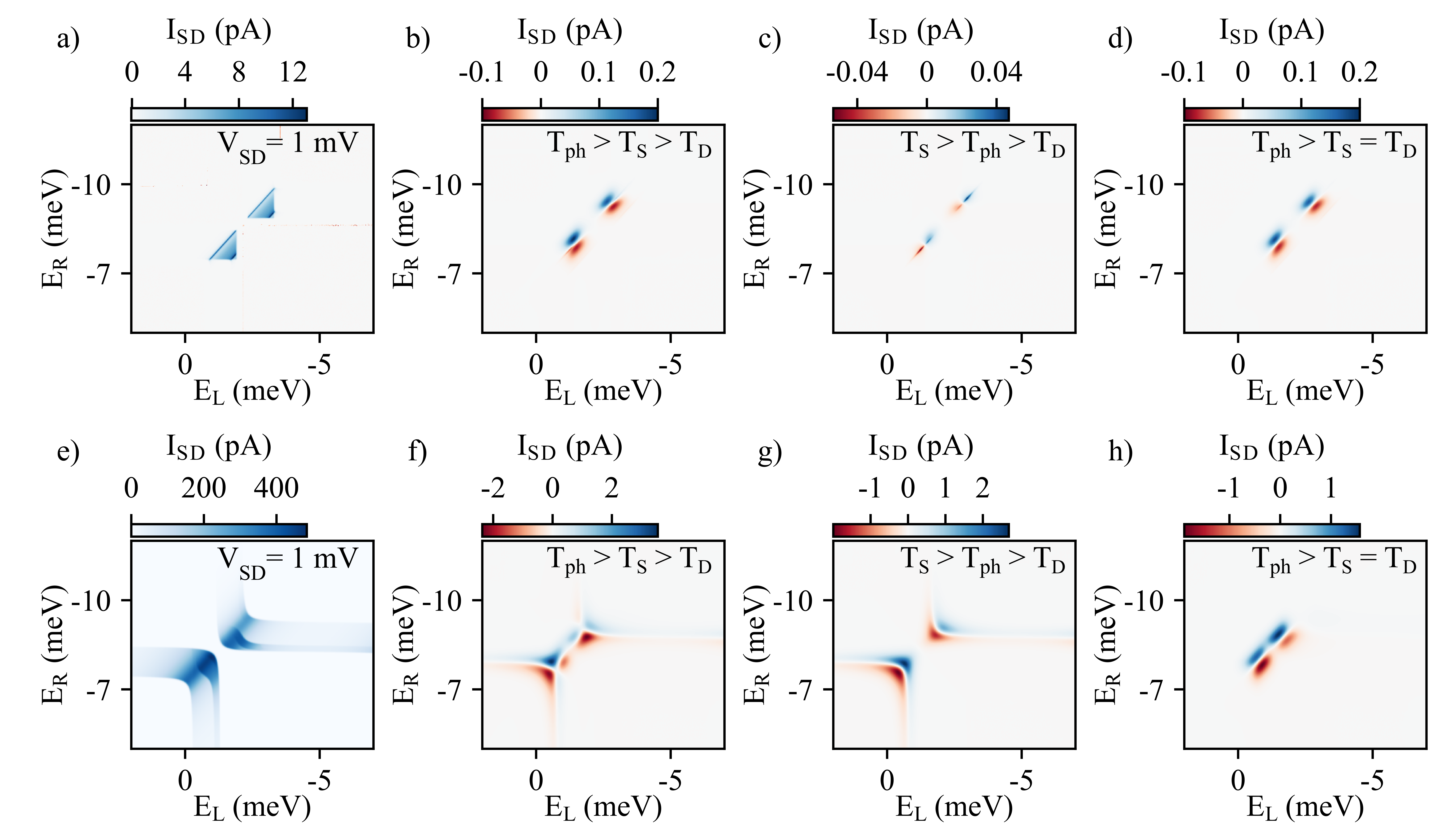}
\caption{Simulated charge stability diagram around the (0,1)$\to$(1,2) charge state transition in the weak (a-d) and strong (e-h) interdot coupling regime. a) and b) show finite bias spectroscopy at $V_{\textrm{SD}}=1\,\si{mV}$, $dV_{\textrm{H}}=0\,\si{V}$ and $T_{\textrm{S}}=T_{\textrm{D}}=T_{\textrm{ph}}=130\,\si{mK}$. Heat driven currents are simulated at $V_{\textrm{SD}}=0\,\si{mV}$ and $dV_{\textrm{H}}=1\,\si{V}$ in the weak (b-d) and strong (f-h) interdot coupling regime. (b,f) showcase the nominal temperatures from eqs.~(\ref{EqTLTRfit},\ref{EqTphfit}). In (c,g)  $T_{\textrm{ph}}$ is replaced by the average of $T_{\textrm{S}}$ and $T_{\textrm{D}}$, while in (d,h) we use $T_{\textrm{S}}=T_{\textrm{D}}=0.9\, \si{K/V}\cdot dV_{\textrm{H}}\,\si{V}+130\,\si{mK}$ and eq. \ref{EqTphfit}.\label{fig_S4}}
\end{figure}

In fig. 3 in the main text and the corresponding discussions, we establish good agreement between the experimental system and the simulations. Thus, in the following, using our simulations, we can disentangle the TE effect and PAT and thereby demonstrate how the experimentally observed signal arises. Figure \ref{fig_S4}(a) and (e) show simulated finite bias spectroscopy around the (0,1)$\to$(1,2) charge state transition at $V_{\textrm{SD}}=1\,\si{mV}$ and $T_{\textrm{ph}}=T_{\textrm{S}}=T_{\textrm{D}}=130\,\si{mK}$ in the weak and strong interdot coupling regime, respectively. The weak interdot coupling regime is dominated by the common finite bias triangles. In contrast, in the strong coupling regime, anti-crossing behaviour is observed near the TPs and charging lines carry current. The first excited state at $0.8\,\si{meV}$ manifests within the finite bias triangles and along the charging lines. Simulations in the weak coupling regime agree well with the experimental data in fig. \ref{fig_S2}(a). Simulations in the strong interdot coupling regime are qualitatively comparable to the experiments in the intermediate interdot coupling regime. 

In fig. \ref{fig_S4}(b,f) we show the current driven by moderate heating ($dV_{\textrm{H}}=1\,\si{V}$) for the weak and strong interdot coupling regime respectively and $T_{\textrm{ph}}>T_{\textrm{S}}>T_{\textrm{D}}$ as defined by eqs.~(\ref{EqTLTRfit},\ref{EqTphfit}). The simulated current patterns agree well with the experiment in the weak interdot coupling regime, where the predominant current polarity reversal manifests along $\Delta$, around the charge transfer line. In the strong interdot coupling regime detailed agreement with the experiment is inferior, but qualitatively the sign changes of the current, observed in experiments (see fig. \ref{fig_S1}(b)), both along and perpendicular to the charge transfer line is captured.

Next, we remove the PAT contribution by reducing $T_{\textrm{ph}}$ to $(T_{\textrm{S}}+T_{\textrm{D}})/2$, such that $T_{\textrm{S}}>T_{\textrm{ph}}>T_{\textrm{D}}$. In this case, see panels (c,g), we recover the thermoelectric current along the charge transfer line which reverses in polarity upon crossing a TP. While this feature is confined to a very narrow range around zero detuning between the ground states in the weak coupling regime (c), it extends along the charging lines for the strong coupling regime (g).

Conversely, the thermoelectric effect is turned off by setting $T_{\textrm{S}}=T_{\textrm{D}}=0.9\,\si{K/V}\cdot dV_{\textrm{H}}+130\,\si{mK}$. As a result, for the case $T_{\textrm{ph}}>T_{\textrm{S}}=T_{\textrm{D}}$, simulated in fig. \ref{fig_S4}(d,h), pure PAT is driving the current. In this case, the signals observed for both the weak and strong interdot coupling are comparable, showing current of opposite polarity exclusively on either side of the charge transfer line in vicinity to the TPs. Weak resonances are present in the weak coupling regime, where $T_{\textrm{ph}}$ provides phonons with sufficiently high energy to transfer electrons into excited states.

By considering the pure thermoelectric and PAT cases in fig. \ref{fig_S4}(c/g) and (d/h), it becomes clear how the complex heat driven signals in fig. \ref{fig_S4}(b/f) emerge by adding the independent thermoelectric and PAT contributions. We note that a comparison of (g) and (h) clearly indicates that currents along the charging lines are of pure thermoelectric nature, giving access to an independent contact electron temperature characterization.

\begin{figure}
\includegraphics[width=8cm]{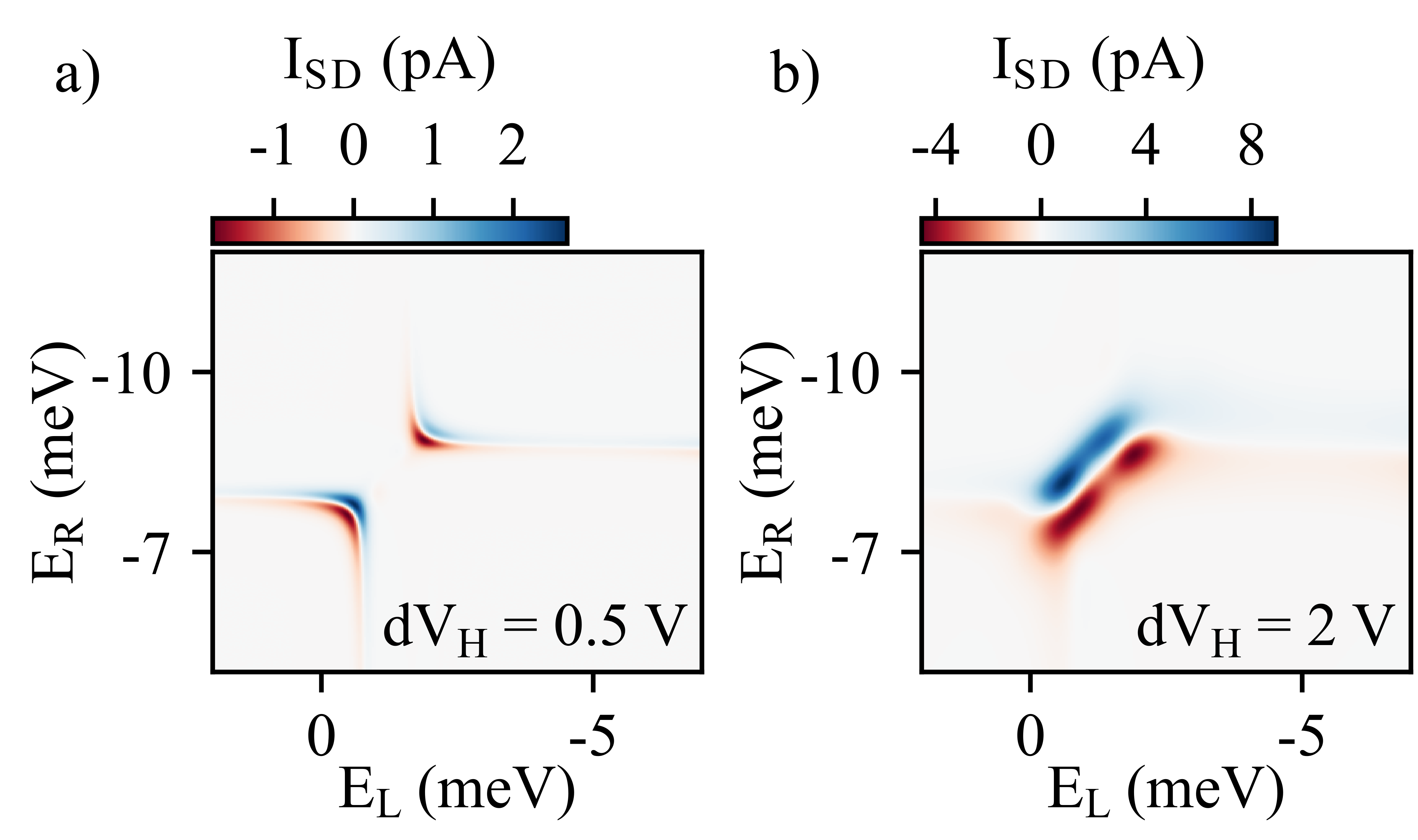} 
\caption{Simulated heat driven currents at $V_{\textrm{SD}}=0\si{V}$ and $T_{\textrm{ph}}>T_{\textrm{S}}>T_{\textrm{D}}$, according to the calibration functions (eqs.~(\ref{EqTLTRfit},\ref{EqTphfit})). a) Simulation with $dV_{\textrm{H}}=0.5\si{V}$ at $k_\textrm{B}T_{\textrm{ph}} < 2\Omega$, resulting in only the thermoelectric current contribution. b) Simulating with $dV_{\textrm{H}}=2\si{V}$, resulting in both, the thermoelectric and PAT current contributions.\label{fig_S5}}
\end{figure}

In the weak interdot coupling regime thermoelectric currents (fig. \ref{fig_S4}(c)) are an order of magnitude smaller than PAT currents (fig. \ref{fig_S4}(d)) for the given temperature gradient. This is a result of the PAT driving temperature difference $T_{\textrm{ph}}-T_{\textrm{S}}=0.3\,\si{K/V}\cdot dV_{\textrm{H}}$ growing faster with $dV_{\textrm{H}}$ compared to the difference between the contact electron temperatures $T_{\textrm{S}}-T_{\textrm{D}}=0.03\,\si{K/V}\cdot dV_{\textrm{H}}$ driving the thermoelectric effect. In the strong interdot coupling regime, we have to bear in mind, that $2\Omega$ provides a minimal energy difference for the splitting between tunnel-coupled states. Thus, for $k_\textrm{B}T_{\textrm{ph}}\ll 2\Omega$, PAT is strongly suppressed as the required phonon absorption becomes unlikely. In the strong interdot coupling case, we have $2\Omega/k_\textrm{B}= 4\,\si{\K}$, explaining the rather weak PAT signal for $dV_{\textrm{H}}=1\, \si{\V}$ and $T_{\textrm{ph}}=1.33\,\si{\K}$.
In contrast, for $dV_{\textrm{H}}=2\, \si{\V}$, we have $T_{\textrm{ph}}=2.53\,\si{\K}$, and PAT is much stronger, becoming the predominant current driving mechanism, see fig. \ref{fig_S5}(b). On the other hand, upon decreasing  $dV_{\textrm{H}}$, the PAT contribution entirely vanishes for the strong interdot coupling case, see fig.~\ref{fig_S5}(a).

\bibliographystyle{achemso}
\bibliography{PAT_SI_Bib}